%% file: main.tex
\documentclass[sigconf]{acmart}

\usepackage{subcaption}
\usepackage{graphicx}      
\usepackage{booktabs} 
\usepackage{xcolor}
\usepackage[utf8]{inputenc}
\usepackage{textgreek}
\usepackage{framed}
\usepackage{listings}
\usepackage{balance}

\definecolor{atomicorange}{RGB}{255,99,71}
\definecolor{cobalt}{RGB}{0,71,171}
\colorlet{cobaltdark}{cobalt!80!black}   

\newcommand{\atomix}{\textsc{AtoMix}}
\newcommand{\baseline}{\textsc{baseline}}

\newcommand{\iconword}[2]{%
  \raisebox{-0.2ex}{\includegraphics[height=1em]{#1}}\ #2%
}

\definecolor{framebg}{RGB}{242,242,242}  
\definecolor{framecolor}{RGB}{64,64,64}  

\newenvironment{graybox}[1]{%
  \def\boxtitle{#1}%
  \MakeFramed{\advance\hsize-\width\FrameRestore}%
  \noindent\textbf{\boxtitle}\par\vspace{4pt}%
}{\endMakeFramed}

\AtBeginDocument{%
  }

\copyrightyear{2026}
\acmYear{2026}
\setcopyright{cc}
\setcctype{by}
\acmConference[C\&C '26]{Creativity and Cognition}{July 13--16, 2026}{London, United Kingdom}
\acmBooktitle{Creativity and Cognition (C\&C '26), July 13--16, 2026, London, United Kingdom}
\acmDOI{10.1145/3803784.3807530}
\acmISBN{979-8-4007-2583-8/2026/07}

\begin{document}

\title{AtoMix: Fostering Structured Visual Ideation for Remote Groups through Atomic Composition and Cross‑Pollination
}


\author{Heeji Jolie Kim}
\affiliation{
  \department{Industrial Design}
  \institution{KAIST}
  \city{Daejeon}
  \country{Republic of Korea}
}
\email{jolieheejikim@kaist.ac.kr}

\author{Anam Ahmad Khan}
\affiliation{
  \department{Industrial Design}
  \institution{KAIST}
  \city{Daejeon}
  \country{Republic of Korea}
}
\email{anam.khan@kaist.ac.kr}

\author{SeungHui Huh}
\affiliation{
  \department{Industrial Design}
  \institution{KAIST}
  \city{Daejeon}
  \country{Republic of Korea}
}
\email{seunghui@kaist.ac.kr}

\author{Andrea Bianchi}
\affiliation{
  \department{Industrial Design}
  \institution{KAIST}
  \city{Daejeon}
  \country{Republic of Korea}
}
\email{andrea@kaist.ac.kr}

\renewcommand{\shortauthors}{Kim et al.}
\newcommand\red[1]{\textcolor{red}{#1}}

\begin{abstract}
  \input{sections/abstract}
\end{abstract}



\maketitle

\section{Introduction}
\input{sections/introduction}

\section{Background and Related Works}
\input{sections/relatedworks}

\section{Formative Study}
\input{sections/formativestudy}

\section{AtoMix}
\input{sections/system}

\section{User Study}
\input{sections/userstudy}

\section{Results}
\input{sections/results}

\section{Discussion}
\input{sections/discussion}

\section{Conclusion}

\input{sections/conclusion}

\begin{acks}
This work was supported by the National Research Foundation (NRF), Korea, under Grant RS-2024-00436931.
\end{acks}

\bibliographystyle{ACM-Reference-Format}  
\bibliography{references}                  


\appendix

\section{Participant Demographics}
\label{appendix:demographics}

\begin{table}[h]
\centering
\small
\renewcommand{\arraystretch}{1.15} 
\begin{tabular}{@{} l p{0.65\linewidth} @{}}
\toprule
\textbf{Gender} & Male: 6,\; Female: 18 \\ \addlinespace[0.6ex]
\textbf{Age} & Mean = 25.4,\; SD = 3.3 \\ \addlinespace[0.6ex]
\textbf{Majors} &
Industrial Design (10) \\
& Spatial Design (5) \\
& Culture Design Management (1) \\ 
& Business Administration (1) \\
& Computer Science and IT Convergence (2) \\
& Graduate School of Culture Technology (2) \\
& Mechanical Engineering (1), Bio-Convergence (1), \\
& School of Transdisciplinary Studies (1) \\ \addlinespace[0.6ex]
\bottomrule
\end{tabular}
\caption{Participant demographics (N=24).}
\label{tab:demographics}
\end{table}

\section{Task 1: Iterative Visual Development (Vertical Continuity)}
\label{appendix:prompt}

For each column of canvases (6 rounds representing sequential contributions to one idea slot), we evaluated visual continuity and idea progression using the following prompt:

\begin{graybox}{Vertical Continuity Evaluation Prompt}
\small
\texttt{You are evaluating visual idea development in a collaborative 6-3-5 brainwriting session. You will see 6 images representing the sequentially developing idea slot (vertically) across 6 sequential rounds, where different participants built upon previous contributions.}

\textbf{Task:} Analyze how the visual idea evolves from Round 1 to Round 6.

\textbf{Important: Handling Empty Canvases}\\
Some canvases may be EMPTY (blank, no content). Empty canvases indicate that a participant did not contribute to that round. When evaluating, consider how gaps affect the overall narrative and whether later rounds successfully recover or build upon available content.

\textbf{Evaluation Criteria:}

\textbf{1. Visual Continuity (1-5)}\\
How well do later rounds maintain recognizable visual elements from earlier rounds?
\begin{itemize}
    \item 1: No connection; each round appears unrelated to others; the visual thread is completely lost
    \item 2: Weak connection; only superficial or accidental similarities; development chain is broken
    \item 3: Moderate connection; some elements carried forward but inconsistently; partial continuity
    \item 4: Strong connection; clear visual thread with intentional variations throughout
    \item 5: Excellent continuity; cohesive evolution with consistent style/elements; seamless progression
\end{itemize}

\textbf{2. Idea Progression (1-5)}\\
Does the concept develop, expand, or deepen across rounds?
\begin{itemize}
    \item 1: No meaningful progression; ideas appear static, random, or disconnected across rounds
    \item 2: Minimal progression; only superficial changes; limited evidence of building upon prior work
    \item 3: Moderate progression; some meaningful development visible, though uneven or interrupted
    \item 4: Strong progression; clear elaboration and refinement; ideas grow in complexity or detail
    \item 5: Excellent progression; systematic building with increasing sophistication; rich cumulative development
\end{itemize}

\textbf{3. Continuity Analysis}\\
Identify any gaps in the development chain:
\begin{itemize}
    \item \texttt{has\_gaps}: true/false
    \item \texttt{gap\_locations}: List of round numbers with empty canvases
    \item \texttt{recovery\_quality}: If gaps exist---``strong'', ``partial'', ``weak'', or ``n/a''
\end{itemize}

\textbf{4. Development Type (categorical)}\\
Classify the dominant pattern: Linear, Branching, Iterative, Discontinuous, Convergent, or Fragmented.

\textbf{Output Format:} JSON with fields for continuity\_analysis, visual\_continuity (score, reasoning), idea\_progression (score, reasoning), development\_type, key\_observations, strongest\_development, and weakest\_development.
\end{graybox}

\section{Task 2: Horizontal Idea Differentiation (Within-Row Diversity)}

For each row of canvases (3 canvases representing distinct ideas from the same round), we evaluated visual distinctiveness and conceptual diversity using the following prompt:

\begin{graybox}{Horizontal Diversity Evaluation Prompt}
\small
\texttt{You are evaluating idea diversity in a brainstorming session. You will see 3 images created by the same participant in the same round. Each image should represent a DIFFERENT idea for the same design challenge.}

\textbf{Task:} Assess how distinct these 3 ideas are from each other.

\textbf{Evaluation Criteria:}

\textbf{1. Visual Distinctiveness (1-5)}\\
How visually different are the 3 canvases?
\begin{itemize}
    \item 1: Nearly identical; same composition/elements with trivial differences
    \item 2: Low distinctiveness; same visual approach with minor variations
    \item 3: Moderate distinctiveness; some shared elements but different arrangements
    \item 4: High distinctiveness; clearly different visual approaches
    \item 5: Maximum distinctiveness; completely different visual styles/compositions
\end{itemize}

\textbf{2. Conceptual Diversity (1-5)}\\
How different are the underlying ideas/concepts?
\begin{itemize}
    \item 1: Same idea expressed 3 times (redundant)
    \item 2: Variations of one core idea (e.g., same feature, different settings)
    \item 3: Related but distinct concepts (e.g., same theme, different features)
    \item 4: Different concepts with some thematic connection
    \item 5: Completely different conceptual approaches
\end{itemize}

\textbf{3. Redundancy Detection}\\
Identify any redundancy:
\begin{itemize}
    \item ``None'': All 3 ideas are meaningfully different
    \item ``Partial'': 2 ideas overlap significantly, 1 is distinct
    \item ``High'': All 3 ideas are essentially the same concept
\end{itemize}

\textbf{Output Format:} JSON with fields for visual\_distinctiveness (score, reasoning), conceptual\_diversity (score, reasoning), redundancy, idea\_summaries (3 brief descriptions), most\_distinct\_pair, and most\_similar\_pair.
\end{graybox}

\end{document}

%% file: sections/abstract.tex
6-3-5 brainwriting is a structured, silent ideation method where 6 participants each generate 3 ideas in 5 minutes across 6 rounds, yielding up to 108 ideas and enabling equal participation, rapid ideation and cumulative development. Many ideas are fundamentally visual, making generative image models especially useful for accelerating visual exploration and making abstract ideas quickly tangible, but existing tools cannot fully support visual composition, where ideas are built from atomic elements and iteratively remixed. A formative study (N=18) showed that even with GenAI embedded in a 6-3-5 structure, participants invested effort in crafting monolithic prompts yet struggled to obtain intended results, build on others’ ideas, or work modularly, leading to redundancy and limited divergence. To address these issues, we present \atomix{}, a collaborative visual interface for remote 6-3-5 brainwriting that supports granularity and cross-pollination through atomic prompting, canvas composition, and reuse of image fragments. In a comparative study (N=24), \atomix{} afforded more fine-grained interaction than the baseline and better supported collaborative visual ideation.

%% file: sections/introduction.tex
Ideation is central to creative work in design. Established frameworks such as brainstorming~\cite{692e1dfc-42e1-3d68-a556-1bc77d50cbd3}, brainwriting~\cite{heslin_better_2009}, and brainsketching~\cite{van_der_lugt_brainsketching_2002} structure how ideas emerge in early innovation. Brainstorming, while popular, suffers from well-documented drawbacks. Evaluation apprehension, production blocking, social loafing, and illusions of productivity all constrain the number of unique ideas relative to nominal (solo) ideation~\cite{heslin_better_2009}. Brainwriting was introduced to address these issues by having participants silently write and exchange ideas, reducing status and dominance effects which supports more reflective engagement with others’ contributions~\cite{heslin_better_2009}. Prior work has extended brainwriting into visual domains: brainsketching uses hand-drawn sketches, rather than speech or text, as the primary medium of thought, with participants drawing individually, rotating positions, and iteratively building on one another’s sketches~\cite{van_der_lugt_brainsketching_2002}. 
Compared to text-based brainwriting, visual approaches provoke new ideas through cycles of 'seeing-making-seeing', described as a "reflective conversation"~\cite{schon1992reflective} with visual materials~\cite{van_der_lugt_brainsketching_2002}. We use the term \textit{visual brainwriting} to describe the process of conducting brainwriting by incorporating visual media instead of merely using written text.

Recent work shows that \emph{AI-augmented brainwriting} can further strengthen early ideation by increasing idea fluency, accelerating elaboration and reframing, and supporting movement between divergence and convergence~\cite{shaer_ai-augmented_2024}. As ideas grow more complex, AI-augmented ideation can begin to break down: generated suggestions often drift from user intent, requiring repeated regeneration and repair~\cite{Peter2025,Jiaxin_2025}. Moreover, while AI can increase idea fluency, it may also promote early convergence by steering participants toward a narrow set of coherent ideas~\cite{holzner2025generativeaicreativitysystematic,Joongi2025}. These tensions raise questions about how AI should be integrated into ideation practices without undermining their core strengths.
Motivated by these findings, we set out to examine whether these limitations of AI-augmented brainwriting also arise in Generative AI (GenAI) supported \emph{visual brainwriting}, specifically image generation, where ideas are developed through iterative interaction with visual materials rather than text alone. To explore this, we first conducted a formative study (N=18) using a digital visual brainwriting board following a 6–3–5 framework augmented with generative image tools to observe how participants use GenAI with a visual brainwriting task. Our results suggests that participants fell into prompt–regenerate loops that treated AI outputs as finished artifacts rather than manipulable materials, and AI-generated images fell short when ideas required multi-component or process-level reasoning. Together, these frictions shifted ideation toward whole-image regeneration and convergence, undermining visual brainwriting’s core strengths of incremental iteration, externalization, and cumulative build-on.

To address these limitations and better realize visual brainwriting’s potential in digital environments, we propose a granular, fragment-level approach grounded in prior evidence that ambiguous, manipulable visual materials support iterative reinterpretation, idea discovery, and incremental build-on connections~\cite{tversky_visualizing_2011,van_der_lugt_brainsketching_2002,suwa_what_1997}. We introduce \atomix{}, a GenAI-augmented visual brainwriting system that re-centers \emph{composition over completion}. \atomix{} provides (i) \emph{collage-based canvas} to generate object-level visuals that drop into the canvas as movable layers; (ii) \emph{region-based editing} that aligns new content to user-selected areas without replacing the whole; and (iii) \emph{cross-canvas mashups} that surface recombination opportunities and carry provenance for traceable build-on. These affordances may help mitigate the drift toward whole-image regeneration and early convergence by enabling atomic, incremental, and cumulative visual ideation, while still leveraging AI generation for speed and fluency.

To explore how these affordances shape visual brainwriting practice, we conducted a comparative study (N=24) examining artifact characteristics and interaction patterns in \atomix{} versus a \baseline{} system. Our study showed that \atomix{} produced canvases with greater compositional richness, higher vertical continuity across rounds, stronger horizontal differentiation between ideas, and fewer empty canvases compared to the \baseline{}. These outcomes were driven by three emergent interaction patterns: atomic prompting (shorter, more frequent prompts), granular editing (region-based modifications), and cross-pollination (reusing and remixing fragments across canvases). Overall, this research makes the following contributions:
\begin{itemize}
    \item \atomix{}, a collaborative visual brainwriting system enabling atomic composition, granular editing, and cross-canvas recombination;
    \item a design approach that positions GenAI outputs as compositional materials rather than complete images, supporting iterative assembly over monolithic generation;
    \item empirical findings from a comparative study (N=24) examining output characteristics (compositional expressiveness, iterative development, idea differentiation), interaction patterns (atomic prompting, granular editing, cross-pollination), and user experience in \atomix{} versus \baseline{} conditions.
\end{itemize}

%% file: sections/relatedworks.tex
We aim to propose a workflow for visual brainwriting that helps designers work more efficiently in group-based ideation practices, using generative AI to enable and accelerate the visualization of ideas. To ground this approach, we review three strands of prior work: (1) \emph{brainwriting and visual brainwriting} as structured, group-oriented ideation methods, (2) \emph{creativity support tools (CSTs)} that scaffold ideation, and (3) \emph{GenAI approaches} that augment ideation, including visualization of ideas and the generation/recombination of concepts.

\subsection{Brainwriting}
Brainwriting addresses traditional brainstorming challenges— production blocking, evaluation apprehension, and social loafing—by enabling participants to generate and exchange ideas in parallel rather than speaking sequentially~\cite{heslin_better_2009,bouschery_ai-augmented_2023,langham_simulating_2020}. A common framework is 6--3--5, where six participants each produce three ideas and pass their sheets every five minutes, ensuring structured rotation and equal contribution~\cite{rombach_high-resolution_2022,shah2001collaborative}. \textit{Visual brainwriting} extends this approach by using visual media as the unit of thought. The original form, brainsketching~\cite{van_der_lugt_brainsketching_2002}, has participants sketch individually, then rotate to build on others' marks, creating incremental "seeing–making–seeing" cycles that promote reinterpretation and cumulative development~\cite{van_der_lugt_brainsketching_2002,tversky_visualizing_2011,suwa_what_1997}. These cycles yield partial, evolving artifacts rather than finished drawings, supporting divergent exploration and stepwise refinement, which are properties we seek to preserve in visual brainwriting. Central to this process is \emph{cross-pollination}: the reuse and recombination of ideas across participants, a form of cross-fertilization where exposure to others' contributions stimulates new associations~\cite{nijstad_production_2002,kohn_collaborative_2011}. In brainwriting, cross-pollination occurs structurally through sheet rotation, enabling participants to build on, reinterpret, or remix prior work. These dynamics support \emph{divergent thinking}—the generation of multiple, varied solutions from a single starting point~\cite{guiford_creativity_1950,runco_divergent_2012}—alongside \emph{combinational creativity}, where novel ideas emerge from blending existing concepts~\cite{boden_creative_2004}. The strict timing further constrains elaborate rendering, encouraging quick, rough visuals that leave interpretive space for others and transform initially separate sheets into an increasingly interconnected pool of ideas~\cite{van_der_lugt_brainsketching_2002,tversky_visualizing_2011,suwa_what_1997}. However, visual brainwriting also surfaces a persistent \textit{visualization gap}: participants often struggle to express intended concepts visually. They may hesitate to draw in front of others (“I can’t draw”), while time-bounded rounds heighten performance pressure and complex, multi-component ideas resist being compressed into a single image~\cite{van_der_lugt_brainsketching_2002,shah2001collaborative,geyer_ideavis_2012}.

\begin{figure}[h]
  \centering
  \includegraphics[width=1\linewidth]{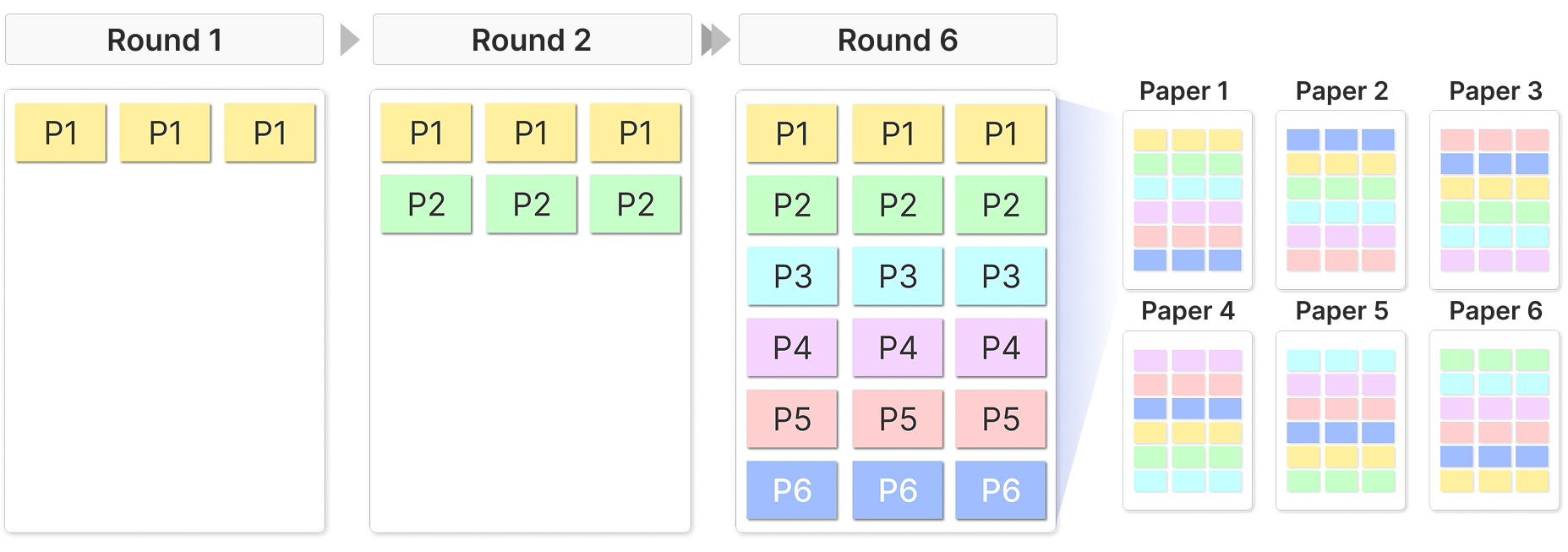}
  \caption{\textbf{Illustration of the 6--3--5 visual brainwriting method}. Six participants (P1--P6) work in parallel; in each 5-minute round, every person creates \emph{three} visual ideas on a sheet and then passes the sheet to the next participant. As rounds progress (Round 1 $\rightarrow$ Round 6), each sheet accumulates layered contributions from all participants. The outcome is six papers, each showing a grid of ideas that reflect incremental build-on~\cite{shah2001collaborative,van_der_lugt_brainsketching_2002}.}
  \label{fig:635}
\end{figure}

This visualization challenge highlights the potential for creativity support tools to lower these barriers while preserving visual brainwriting's key benefits: incremental idea development, collaborative building across participants, and the interpretive flexibility of rough, partial visuals. Our work aims to address the visualization gap through the design of a creativity support tool for visual brainwriting.

\subsection{CSTs for Ideation}

Creativity support tools (CSTs) are interactive systems designed to enhance human creativity across a range of activities, with one important strand focusing on scaffolding ideation through exploration, expression, and problem-solving~\cite{shneiderman_creativity_2007,frich_mapping_2019}. Shneiderman's framework identifies core principles for effective creativity support—low thresholds, wide walls, and high ceilings—alongside the importance of exploration, collaboration, and history-keeping~\cite{shneiderman_creativity_2007}. Many CSTs also build on \emph{direct manipulation}~\cite{shneiderman_direct_1983}, where continuous, reversible operations on visible objects reduce cognitive overhead and invite experimentation. Below, we contrast two strands of CSTs in ideation---\emph{individual} and \emph{collaborative} tools---to identify gaps our work addresses.

\subsubsection{Individual Ideation Support}
CSTs for individual ideation embed context-sensitive cues, domain-specific examples~\cite{cao_idea_2015,jin_creativesvg_2021}, and mechanisms for \emph{curated surprise}: graph-based tools expose non-obvious links~\cite{Bae2020Spinneret}, pen+touch diagramming maintains productive ambiguity~\cite{Webb2015PenTouchDiagramming}, and semantic retrieval steers users toward diverse ideas~\cite{siangliulue_ideahound_2016}. However, most systems operate at the level of \emph{complete} examples or query results rather than manipulable parts, offering limited support for modular recombination or cumulative build-on~\cite{van_der_lugt_brainsketching_2002,tversky_visualizing_2011,suwa_what_1997}. As a result, how individually expanded idea spaces translate into shared, incremental ideation remains underexplored~\cite{shneiderman_creativity_2007,frich_mapping_2019}.

\subsubsection{Collaborative Ideation Support}
Collaborative CSTs strengthen group brainstorming through coordination, visibility, and real-time artifact manipulation. Tangible approaches increase social presence during remote sessions~\cite{melniczuk_exploring_2025} and structure turn-taking~\cite{10.1145/3623509.3633361}. Sketch-centric tools support co-located brainstorming~\cite{kim_tablaction_2010}, moderated remote sketching~\cite{lee2017all4one}, and hybrid workspaces linking paper with interactive layers~\cite{geyer_ideavis_2012}. Group-oriented platforms provide shared canvases with branching and merging~\cite{zhao_skwiki_2014,kazi_sandcanvas_2011}, structured ideation grids~\cite{Knight2021IdeationGrids}, game-based interfaces~\cite{Huang2007IdeaStormingCube}, and semantic modeling~\cite{siangliulue_ideahound_2016}. These systems suggest that collaboration improves when participants can interact rapidly and incrementally with partial artifacts that others can see and extend.

\paragraph{\textbf{Takeaways on CSTs for Ideation}}
Individual CSTs expand personal idea spaces; collaborative platforms make thinking visible in shared workspaces~\cite{shneiderman_creativity_2007,frich_mapping_2019}. Despite the availability of direct manipulation~\cite{shneiderman_direct_1983} and creativity support principles~\cite{shneiderman_creativity_2007}, most tools still treat ideas as complete products rather than \emph{manipulable materials} that can be broken apart, modified, and reused—the very mechanics that sustain cumulative build-on in visual brainwriting~\cite{van_der_lugt_brainsketching_2002,tversky_visualizing_2011,suwa_what_1997}.

\subsection{GenAI for Early-Stage Ideation}

GenAI has lowered barriers to visual exploration through text-to-image systems like ChatGPT\footnote{\url{https://chatgpt.com/}}, Stable Diffusion\footnote{\url{https://stablediffusionweb.com/}}, and Midjourney\footnote{\url{https://www.midjourney.com/}}, enabling rapid generation of intended or unexpected visuals that stimulate divergent thinking~\cite{rombach_high-resolution_2022,saharia_photorealistic_2022}.

\subsubsection{GenAI for Idea Generation}
GenAI-for-ideation splits into two trajectories. One embeds AI into \emph{group formats}, extending brainwriting with models that propose, cluster, or evaluate ideas~\cite{shaer_ai-augmented_2024,zhang_ladica_2025}. Another targets \emph{designer control} through iterative exploration, reference reuse, and prompt scaffolds~\cite{wang_aideation_2025,kwon_designer-generative_2024}. Adjacent work positions AI as a \emph{sketching partner}: co-creative tools return complementary sketches using analogy to widen the search space~\cite{kim_collaborative_2021,zhang_generative_2023,lin_inkspire_2025}, while domain tools leverage mood boards and concept blending~\cite{choi_creativeconnect_2024, sun_creative_2025,jeon_fashionq_2021,wan_gancollage_2023}. However, these systems typically treat ideas as \emph{whole renders} or \emph{keyword bundles}, with recombination remaining a single-user activity~\cite{heslin_better_2009,tversky_visualizing_2011,suwa_what_1997}. This monolithic interaction encourages regenerate-and-replace workflows, limiting shared buildup across people and rounds.

\subsubsection{GenAI for Visualization}
Generative models increasingly turn abstract concepts into visual forms. Within this space, one strand streamlines prompt interaction through tools that decompose and blend concepts or expand the design space via prompt suggestion~\cite{wang_popblends_2023,10.1145/3613904.3642847,10.1145/3582269.3615596}, while another emphasizes recombination, blending image elements or integrating generation with reference retrieval~\cite{choi_creativeconnect_2024,sun_creative_2025,wang_aideation_2025}. Parallel to these research tools, modular generative workflows such as ComfyUI\footnote{\url{https://www.comfy.org/}} expose generation pipelines as node graphs, giving users fine-grained control over individual processing steps; however, these environments target technical users and do not support real-time collaborative ideation. Building on these directions, collaborative tools further combine mood boards with semantic search and biological analogies~\cite{koch_imagesense_2020,kang_biospark_2024,haddad2025user,kwon_designer-generative_2024}.

\paragraph{\textbf{Takeaways on GenAI for Ideation}}
What remains under- explored is not \emph{whether} AI can produce visuals, but \emph{how} those visuals function as manipulable materials in collaborative ideation. Current systems return \emph{complete renders}—effective for seeding ideas but inadequate for visual brainwriting's mechanics: (i) incremental composition of partial visuals rather than finished images; (ii) situated extension through in-place transformation rather than regeneration; and (iii) idea circulation where visible lineage sustains reinterpretation across participants~\cite{van_der_lugt_brainsketching_2002,tversky_visualizing_2011,suwa_what_1997}. Current GenAI is optimized for \emph{finished images}, whereas visual brainwriting gains power from \emph{manipulating elements} across time and collaborators.

%% file: sections/formativestudy.tex
To examine the challenges and opportunities of visual brainwriting with generative images, we conducted a formative study using a digital adaptation of the 6–3–5 brainwriting method \cite{wilson_chapter_2013}. This adaptation allowed us to observe how participants used visual content and generative tools during sequential ideation, informing the design of systems for multi-round idea development.

\subsection{Participants and Procedure}

We recruited 18 participants (9 female, 9 male), aged 20–27 (M = 22.4, SD = 2.52), from the institutional community via online postings. Participants represented diverse academic backgrounds, including industrial design, electrical engineering, computer science, and chemistry.

\begin{figure}[h]
  \centering
  \includegraphics[width=1\linewidth]{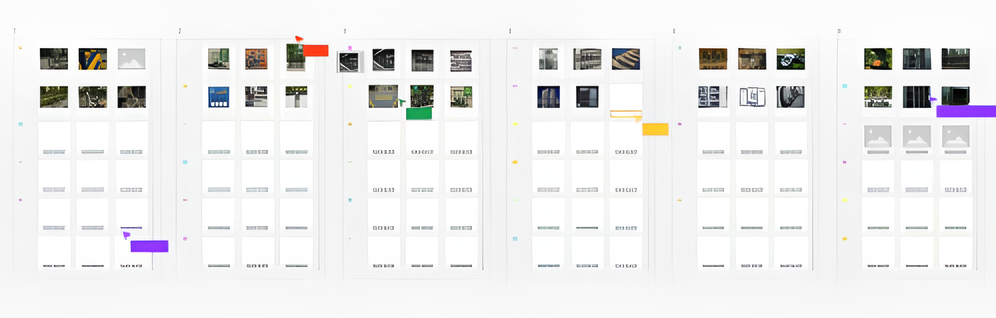}
  \caption{Figma workspace used for formative study--- 6 papers with 6 rows (rounds), 3 columns (ideas), currently in round 2 }
  \label{fig:formativeworkspace}
\end{figure}

The brainwriting activity was conducted in a Figma\footnote{\url{https://www.figma.com/}} workspace replicating the 6–3–5 method and adapted for collaborative visual ideation with generative AI. As shown in Fig.~\ref{fig:formativeworkspace}, the workspace comprised six pages, each with six rows and three columns. To support deeper exploration, we extended each round from 5 to 10 minutes. Participants could use any text or image generative AI tools, including ChatGPT\footnote{\url{https://chatgpt.com/}}
, Gemini\footnote{\url{https://gemini.google.com/}}
, and Copilot\footnote{\url{https://copilot.microsoft.com/}}

\begin{figure}[h]
  \centering
  \includegraphics[width=1\linewidth]{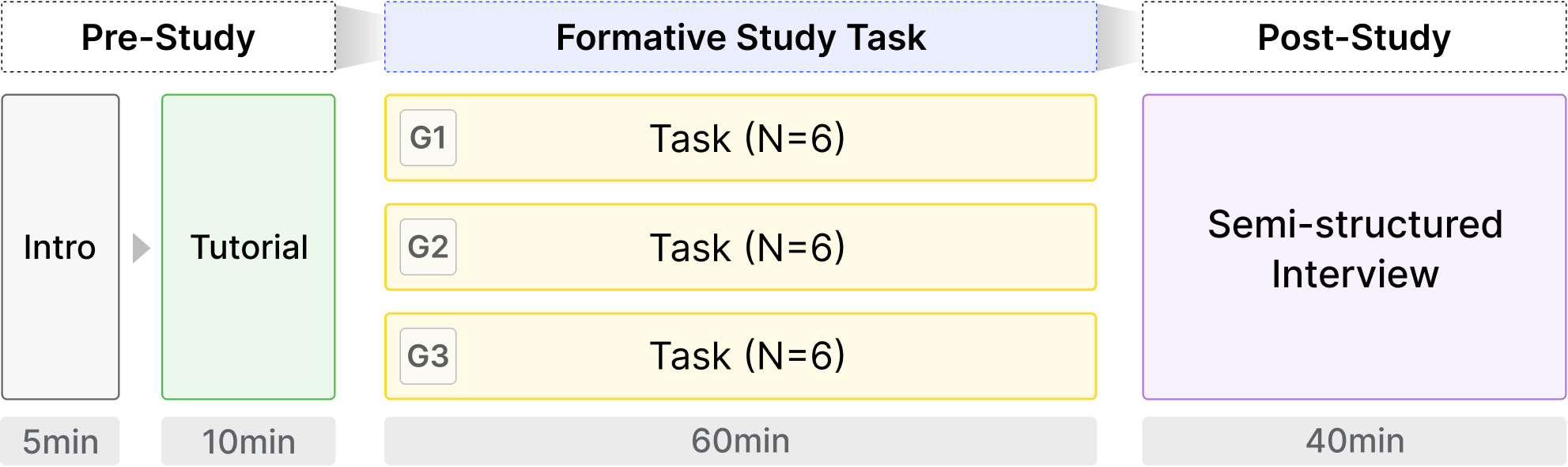}
  \caption{Formative study procedure. Each session comprised an 15-minute introduction (briefing, consent, and tutorial), a 60-minute ideation activity comprised of 6 rounds of 10 minutes, and a 40-minute semi-structured interview on participants’ experiences.}
  \label{fig:formativeprocedure}
\end{figure}

Figure~\ref{fig:formativeprocedure} shows the study procedure. After a brief introduction, participants accessed the Figma workspace and completed the same task: \textit{Design a storage or parking solution for personal micro-mobility devices integrated into urban residential environments}. Using images and text, participants followed the 6–3–5 structure, passing ideas to the next participant for further development. We conducted three 2-hour sessions with six participants each. Each session included six 10-minute rounds, followed by a 40-minute semi-structured group interview exploring participants’ experiences, use of AI tools, and strategies for generating and evolving visual ideas. Interviews were audio recorded, transcribed, and analyzed using thematic analysis~\cite{braun_using_2006}.

\subsection{Findings}
Our analysis showed that GenAI both supported and constrained creativity in visual brainwriting. While participants valued its speed and generative capacity, they encountered intent–output mismatches, limited control, and difficulty expressing complex ideas with single images. Participants adapted by reusing, recombining, and diverging ideas, exposing gaps in existing tools and informing our design goals for balancing AI assistance with creative agency.

\subsubsection{\textbf{Modular and Collage-Based Expression}}
Participants valued AI’s ability to rapidly visualize product ideas, making them easy to grasp at a glance. For instance, P8 noted that tangible concepts like a foldable kickboard were “instantly understandable.” However, AI-generated images were less effective for multi-component or process-oriented ideas (P10, P11). As P10 explained, complex systems require step-by-step representation, and “trying to explain the whole process with just a single picture” often caused confusion. To compensate, participants adopted collage-like strategies, reusing prior outputs as building blocks (P11) or manually recombining images via screenshots in Figma (P12). Many also wanted finer editing control, such as cropping or enlarging specific regions to highlight key elements (P9, P12, P17). Together, these findings show that when single images were insufficient, participants decomposed, reused, and recombined visual elements, motivating the need for compositional expressiveness that support modular editing, layering, and incremental refinement for communicating complex ideas.

\subsubsection{\textbf{Limits of Control and Contextual Refinement}}

Participants valued generative AI for sparking unexpected ideas, especially in early rounds. As P8 noted, “It gave me things I wouldn’t have thought of,” highlighting AI’s role in expanding creative possibilities. However, outputs often misaligned with user intent: refinements produced near-identical images despite prompt changes (P8, P10, P12) or introduced missing or irrelevant elements (P9, P12). Participants tried workarounds such as re-uploading images, simplifying prompts, or wishing for low-fidelity previews, which underscored the lack of fine-grained control. As P17 explained, it would be “much [more] efficient to make small manual edits rather than regenerate images from scratch”. Overall, these findings show that while AI supported inspiration and variation, limited control hindered intent alignment, motivating the need for localized refinement for selective modification.

\subsubsection{\textbf{Creativity Relies on Reuse and Divergence}}
Participants sustained creativity through two complementary strategies: reusing earlier ideas and seeking divergent stimuli when momentum slowed. Reuse involved revisiting prior prompts or images, extracting useful elements, and recombining them to extend ideas without starting over (P2, P14–16). For example, P2 compiled prompts to identify “common elements and improvement points,” while P14 remixed earlier images with new ones. However, participants lacked modular control; as P18 noted, “I wanted to just grab one part of an image, but there’s no easy way to do that,” pointing to needs for cropping, style extraction (P11), preset backgrounds, or shared libraries (P12). When ideation slowed, participants adopted divergent strategies to regain momentum, such as revisiting earlier canvases (P13, P16), feeding accumulated text into ChatGPT (P17), or switching AI tools when outputs became repetitive (P13, P17). 
Together, these findings show that reuse supported continuity while divergence enabled novelty, but the lack of fragment-level editing and structured recombination forced ad hoc workarounds, motivating systems that better support modular reuse, flexible recombination, and divergent exploration across rounds.

\subsection{Design Goals}

Based on insights from our formative study, we identified three design goals to guide the development of our collaborative brainwriting system.
\begin{itemize}
    \item \textbf{DG1: Compositional Expression.} The system should allow users to express complex, multi-component ideas by composing visual elements modularly—selecting, layering, and reusing fragments across rounds—rather than relying on single monolithic images.
    \item \textbf{DG2: Localized Refinement.} The system should support fine-grained modification of specific regions or elements, preserving surrounding context and creative intent without requiring full regeneration.
    \item \textbf{DG3: Recombinational Flow.} The system should sustain ideation momentum by enabling recombination of visual and textual fragments across canvases, supporting both reuse and divergence.\end{itemize}

%% file: sections/system.tex
Guided by the derived design goals, we implemented \atomix{}, a web-based collaborative ideation system that integrates image generation, collage composition, granular editing, and cross-canvas mashup across multiple canvases. We first present a system overview, followed by a description of each feature aligned with the three design goals.

\begin{figure*}[t]
  \centering
  \includegraphics[width=\textwidth]{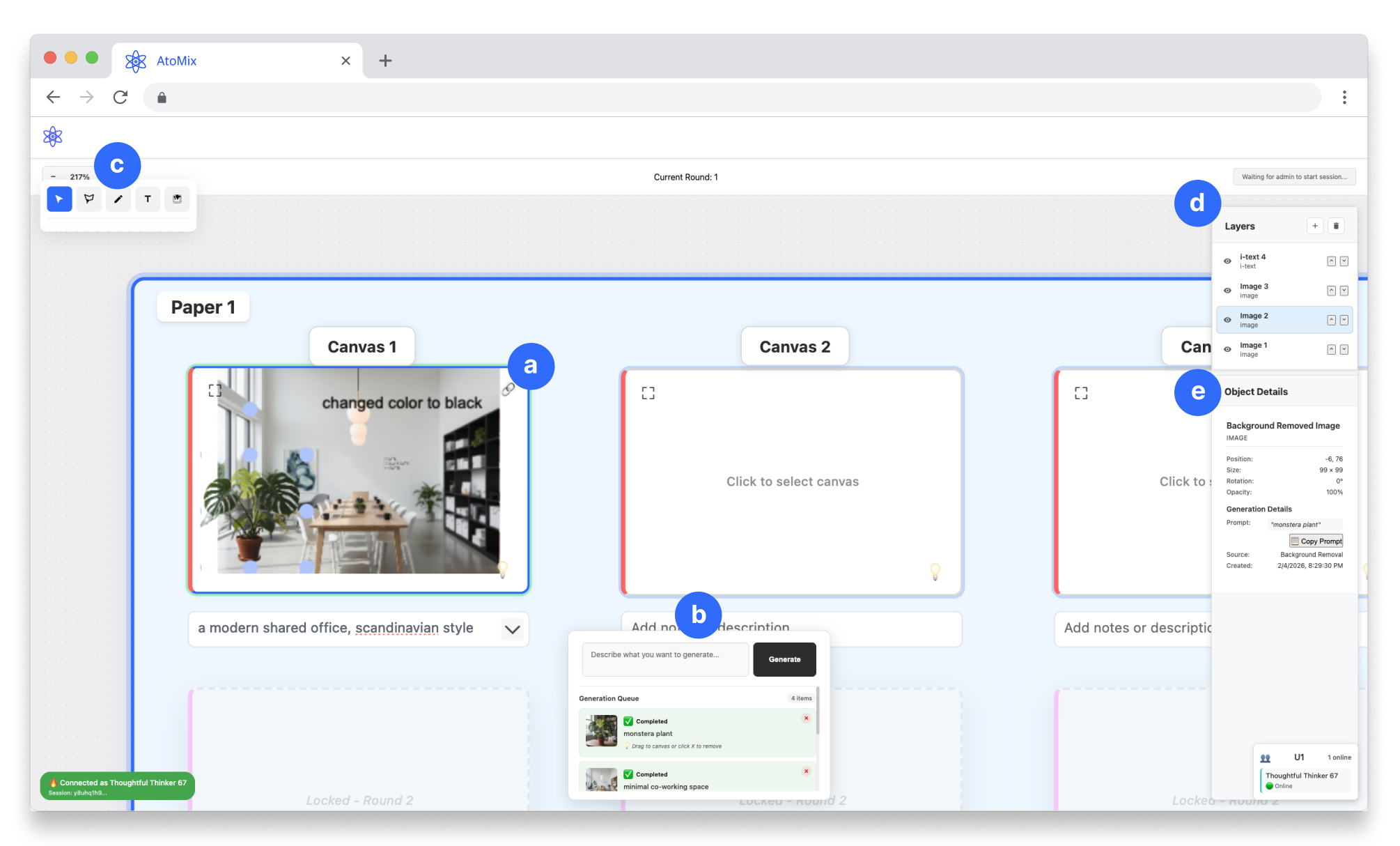} 
    \caption{\textbf{System Screenshot} (a) \textbf{Multi-canvas collage board:} three side-by-side canvases per round; users arrange fragments and add per-canvas notes. (b) \textbf{Prompt \& Generation Queue:} enter a description to generate images; completed results appear with status and can be dragged onto any canvas. (c) \textbf{Toolbar:} selection, lasso for region-based edits, background removal, pen, and text. (d) \textbf{Layers panel:} every fragment is a layer; users reorder, hide, and select layers for precise control. (e) \textbf{Object Details:} per-fragment metadata (prompt, source, timestamp) for provenance and copy-back of prompts. The example depicts a Scandinavian design for a shared office, featuring a plant image assembled from generated fragments and annotations.}
    \label{fig:system}
\end{figure*}

\subsection{System Workflow}

In \atomix{}, we used the 6–3–5 method where participants work through six rounds, with three canvases unlocked per round for five minutes before rotating to the next person's work. Within this structure, users visualize ideas by composing generated images, making granular changes, and synthesizing ideas from multiple canvases.

\begin{figure*}[t]
  \centering
  \includegraphics[width=\textwidth]{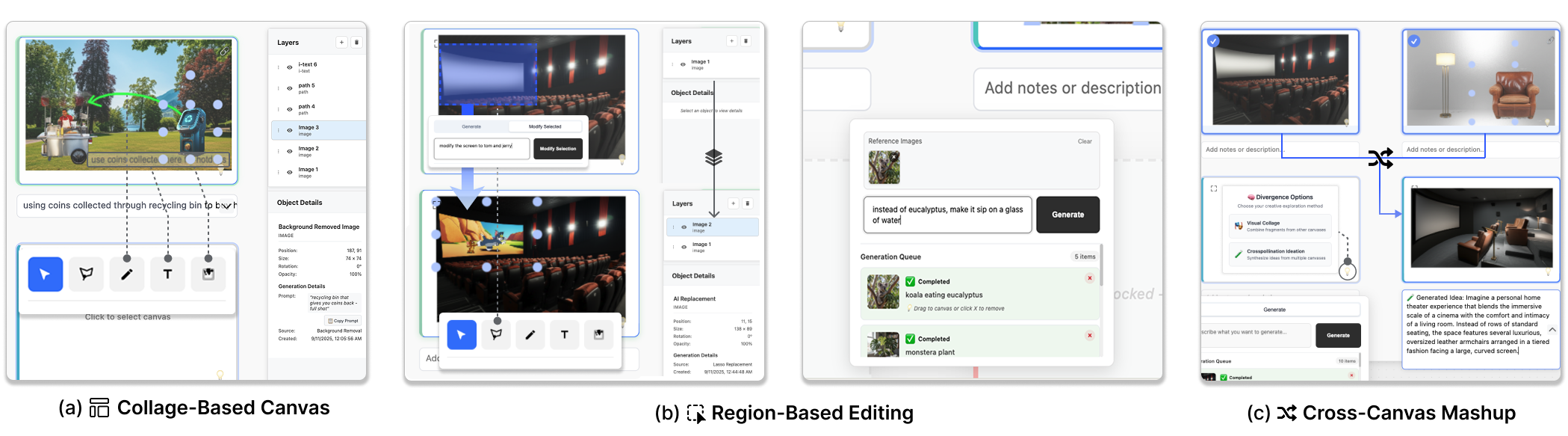} 
    \caption{\textbf{System features.} (a) Collage-Based Canvas—compose AI-generated/imported fragments. (b) Region-Based Editing—re-prompt and refine selected regions in place as a new fragment. (c) Cross-Canvas Mashup— recombine ideas across canvases through reuse or image mashup feature.}  
    \label{fig:features}
\end{figure*}

\subsubsection{\iconword{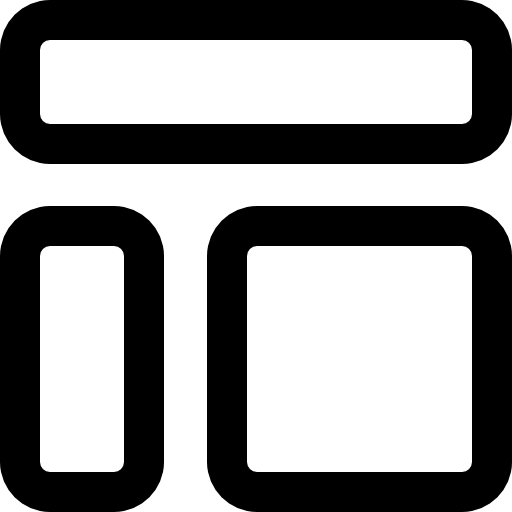}{\textbf{Collage‑Based Canvas}}}

Each canvas functions as a freeform collage space where users generate images by typing descriptions into the prompt panel. Generated images appear in a queue; users drag them onto the canvas and manipulate them as independent fragments—resizing, rotating, repositioning, or deleting. The toolbar provides text boxes, a pen tool, and background removal to support layering images without bounding-box overlap. Each element is stacked on a separate layer with individual visibility control, enabling users to compose multiple images, annotations, and sketches into unified concepts (Figure~\ref{fig:features}a).

\subsubsection{\iconword{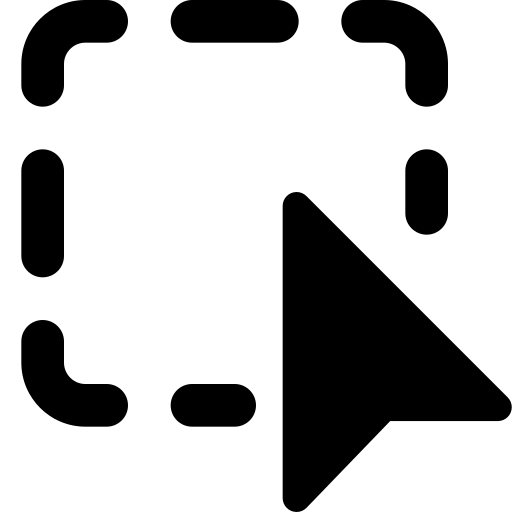}{\textbf{Region‑Based Editing}}}

Region-based editing enables fine-grained modifications without regenerating entire images. Users select the lasso tool, trace a region on an existing image, and enter a short instruction—optionally attaching a reference image. The system sends the selected patch with its local context to the generative model; the returned result is inserted as a new fragment layer aligned to the selected region, leaving the original untouched. A layer panel lets users reorder, hide, or duplicate fragments to compare alternatives (Figure~\ref{fig:features}b).

\subsubsection{\iconword{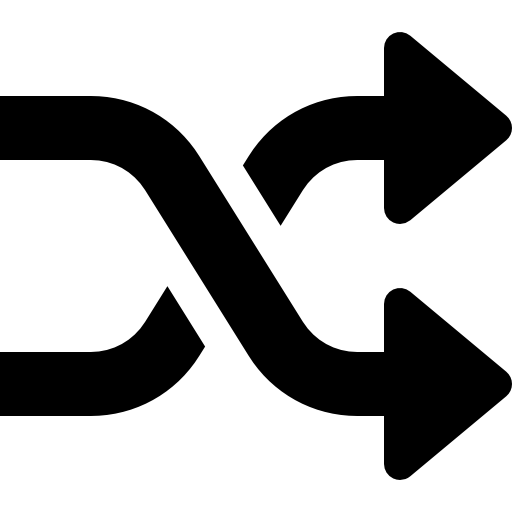}{\textbf{Cross‑Canvas Mashups}}}

By clicking on the "lightbulb" button, users can select two to four canvases and request a mashup. The system extracts snapshots and metadata, blending elements from each source. The generative model produces a new image with a textual description of the combined concept, which users insert onto their current canvas (Figure~\ref{fig:features}c).

Across all features, \atomix{} emphasizes lightweight, compositional workflows that align with visual brainwriting's ethos of ``thinking while doing''~\cite{van_der_lugt_brainsketching_2002}.

\subsection{Implementation}

\atomix{} is a multi-user web application built with JavaScript (ES6) and Fabric.js\footnote{\url{http://fabricjs.com/}}, which provides the object model for the canvas. Each image, sketch, and annotation is treated as an independent object that can be directly manipulated rather than flattened into a bitmap, with metadata storing its originating canvas, creation time, and source prompt.

Firebase Firestore\footnote{\url{https://firebase.google.com/docs/firestore}} maintains the state of papers, canvases, and fragments across participants, with updates synchronized in real time using incremental patches. Lineage metadata is recorded to trace origin across rounds. Generated images are stored on Cloudinary\footnote{\url{https://cloudinary.com/home}}, with URLs stored in Firestore for retrieval in subsequent rounds.

Image generation uses the Gemini 2.5 Flash API\footnote{\url{https://ai.google.dev/gemini-api/docs}}. The system wraps user prompts with uniform settings (resolution, aspect ratio, style) and augments requests with contextual metadata—for region-based edits, this includes the selected patch and surrounding context; for cross-canvas mashup, rasterized snapshots and metadata from multiple canvases are merged to guide synthesis.

%% file: sections/userstudy.tex
The purpose of this study is to explore how our proposed visual brainwriting system influences idea generation and development, and whether its design features lead participants to engage in ideation differently. We conducted a between-subjects study with 24 participants, split into two groups using \atomix{} and two groups using a baseline system. We analyzed emergent ideation patterns during system use and assessed participants’ overall user experience.

\begin{itemize}
\item \textbf{RQ1}: How do ideation artifacts differ between \atomix{} and the \baseline{} system?
\item \textbf{RQ2}: How do interaction patterns in \atomix{} and the \baseline{} system influence ideation artifacts?
\item \textbf{RQ3}: How do participants perceive and experience \atomix{} for visual brainwriting?
\end{itemize}

\subsection{Participants and Procedure}
We recruited 24 participants (18 female, 6 male) via online postings at our institution, nearby universities, and industry and research institutes. Participants had a mean age of 25.4 years (SD = 3.3) and came from diverse academic backgrounds, primarily design, as well as business, engineering, computing, and interdisciplinary programs. All reported prior design or HCI-related experience. Detailed demographics are shown in the Appendix~(\S\ref{appendix:demographics}). 

\begin{figure}[h]
  \centering
  \includegraphics[width=1\linewidth]{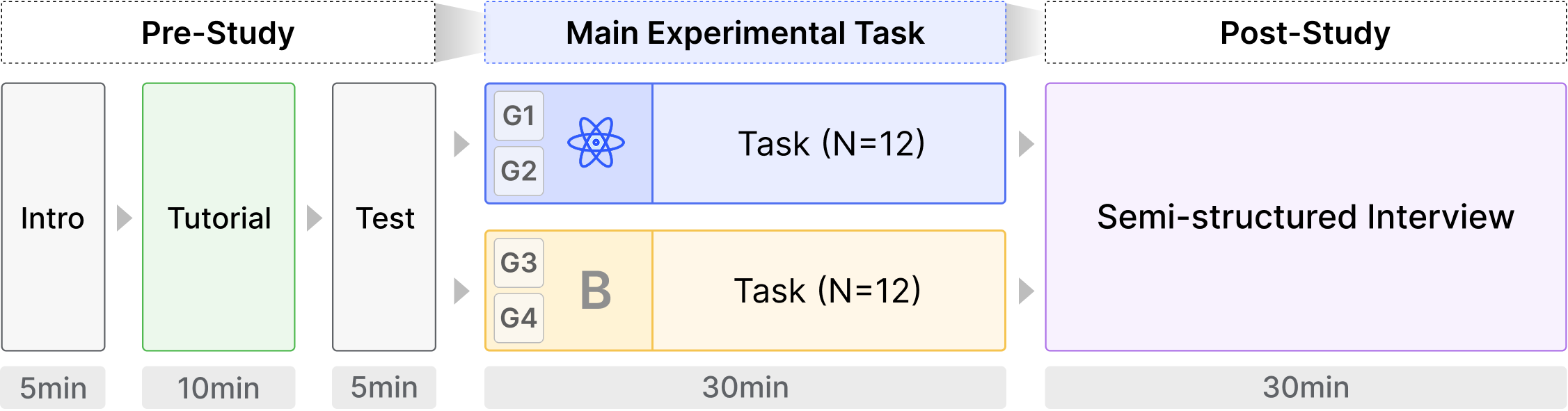}
  \caption{User study protocol. The remote study consisted of a 30-minute ideation phase using either the \baseline{} system or \atomix{}, followed by a 30-minute semi-structured group interview. }
  \label{fig:procedure}
\end{figure}

The study procedure is shown in Fig.~\ref{fig:procedure}. The study was conducted remotely, with participants joining via video conferencing and using the web-based system on personal devices, reflecting real-world remote collaboration. Participants completed a visual brainwriting task using either \atomix{} or a baseline system on the topic \textit{“Design experiences for a learning environment in virtual reality”}. The study sessions lasted 1.5 hours, and participants were compensated approximately \$30 USD.

To isolate \atomix{}’s design effects—integrated generation, granular editing, and cross-canvas recombination—we developed a \baseline{} system that retained the core visual brainwriting structure while omitting these features. Both systems shared the same multi-canvas setup, user assignment, and 6–3–5 round-based automation, ensuring differences stemmed from interaction design rather than collaboration mechanics. \baseline{} participants used a separate image generator with the same model (Gemini 2.5 Flash Image), generating complete images that were copied onto their canvases, reflecting a conventional GenAI-assisted workflow. In contrast, \atomix{} participants completed the same six-round session directly within the deployed web application.

For each group, participants first completed a 10-minute tutorial and a 5-minute practice round to familiarize themselves with system functions (i.e., image generation and modification, collaging, and layer manipulation). They then ideated for 30 minutes on the given topic using the 6–3–5 brainwriting framework, generating three initial ideas and passing their canvas every 5 minutes for six rounds, resulting in a cumulative set of ideas across participants.

\subsection{Measures}
After the task, participants took part in a 30-minute semi-structured \iconword{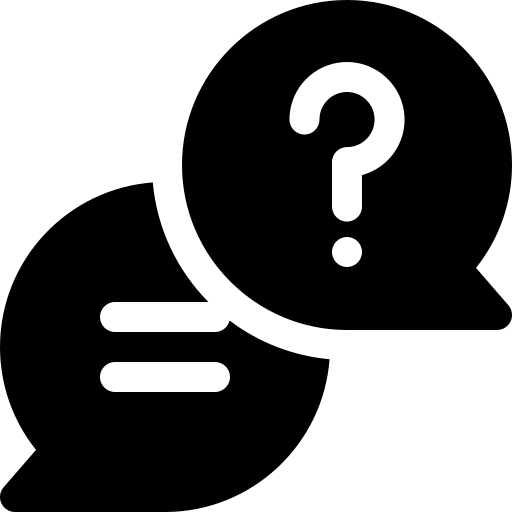} \textbf{group interview} to reflect on their experience with the assigned system, including how it shaped their brainwriting flow and how features such as collaging, image modification, and cross-canvas mashup influenced ideation. We also collected \iconword{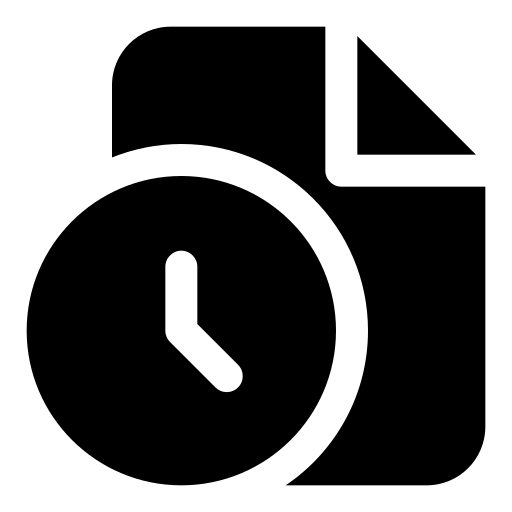} \textbf{user logs} capturing all canvas interactions (i.e., prompting, image editing, tool use, and canvas mashup) with timestamps, as well as prompt submissions and generated outputs. For the \atomix{} condition, we additionally logged region-based edits, fragment reuse, and cross-canvas mashup requests. Screen recordings provided further interaction context.
To assess output characteristics, we collected \iconword{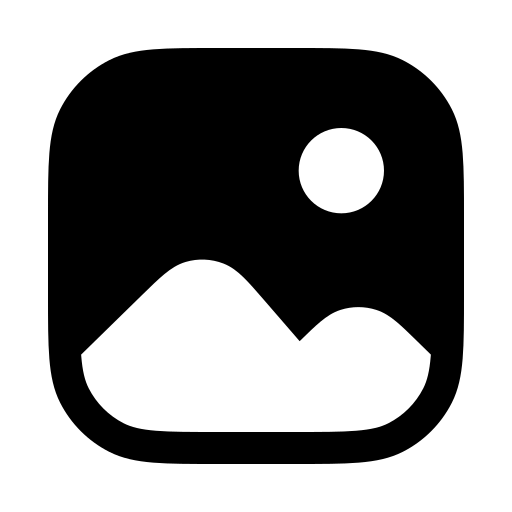} \textbf{canvas visual outputs}, consisting of the final artifacts from each session. Each session produced 108 canvases (6 papers $\times$ 6 rounds $\times$ 3 idea slots), enabling analysis of compositional expressiveness, iterative visual development across rounds, and idea differentiation within rows.

\subsection{Analysis}

\subsubsection{Action Log Analysis}
We logged user interactions at the event level and defined five action categories: \textit{Prompting}, \textit{Tool}, \textit{Text Annotation}, \textit{Refine} and \textit{Cross-pollination}. \textit{Prompting} captured text inputs used to generate images (via the integrated generator in \atomix{} or an external interface in Baseline). \textit{Tool} included non-generative edits such as background removal, drawing, text annotation, lasso-based selection and movement, and layer reordering. \textit{Text Annotation} recorded text description updates beneath each canvases. Two categories were exclusive to \atomix{}. \textit{Refine} captured region-based AI edits using lasso selection and logged an associated prompting event. \textit{Cross-pollination} captured cross-canvas reuse, including copying images from others’ canvases and synthesizing new images from multiple sources. For each event, we recorded participant ID, round, and condition, and aggregated counts by participant, round, and system for analysis.

\subsubsection{LLM-as-Judge}
We used a vision-enabled Large Language Model (LLM) as an automated judge to evaluate visual ideation outputs at scale, enabling consistent, fine-grained assessment \cite{pan2024humancentereddesignrecommendationsllmasajudge,10.1145/3613904.3642216}. We employed Claude Sonnet 4.5\footnote{\url{https://www.anthropic.com/claude/sonnet}} to assess two dimensions of ideation output. First, \textit{iterative visual development} captured how ideas evolved across rounds within each column (6 canvases representing sequential contributions to one idea slot). The LLM rated visual continuity (maintenance of recognizable elements across rounds) and idea progression (development, expansion, or deepening of concepts) on 5-point scales and classified the dominant development pattern (i.e., linear, branching, iterative, discontinuous). Second, \textit{horizontal idea differentiation} captured diversity within each row (3 canvases representing distinct ideas from the same round). The LLM rated visual distinctiveness and conceptual diversity on 5-point scales and flagged redundancy when ideas lacked meaningful differentiation~(\S\ref{appendix:prompt}).

For each evaluation, we used structured prompts, anchored rating scales with examples for each score level, and JSON-formatted outputs, and were blind to system condition. A Python script automated image submission, response parsing, and score aggregation. We analyzed 72 vertical development cases (3 columns × 6 papers × 4 groups) and 144 horizontal differentiation cases (6 rows × 6 papers × 4 groups). To validate LLM judgments, one researcher re-coded a random 20\% subset using the same rubrics, yielding high agreement (92.7\% for visual continuity; 89.2\% for conceptual diversity), consistent with standard reliability thresholds. We acknowledge that LLM-based evaluation may carry systematic biases that a single-coder spot check cannot fully detect; we therefore treat these scores as complementary to, rather than a replacement for, our qualitative and log-based analyses.

\subsubsection{Thematic Analysis}
To examine user perceptions of \atomix{} (RQ3), we analyzed interview transcripts using a general inductive approach \cite{thomas_general_2006}. Two researchers independently coded the data, focusing on perceived strengths and limitations of \atomix{} for brainwriting. One coder proposed 11 initial themes, which were refined by the second coder. Both then reapplied the finalized codes, achieving strong inter-rater reliability (0.80) \cite{mchugh_interrater_2012}. Discrepancies were resolved through discussion, and the findings were organized into themes, highlighting key insights into user experiences with \atomix{}.

%% file: sections/results.tex

\subsection{RQ1. How do ideation artifacts differ between \atomix{} and the \baseline{} system?}
We examined the visual ideation artifacts produced during the brainwriting activity, focusing on how system affordances shaped their form and evolution. We analyze artifact-level characteristics that capture how participants externalized, developed, and differentiated ideas across rounds. Our analysis integrates qualitative observations, interaction logs, and visual model–based evaluations, revealing four key patterns that distinguish artifacts produced with each system.

\begin{figure*}[t]
  \centering
  \includegraphics[width=\textwidth]{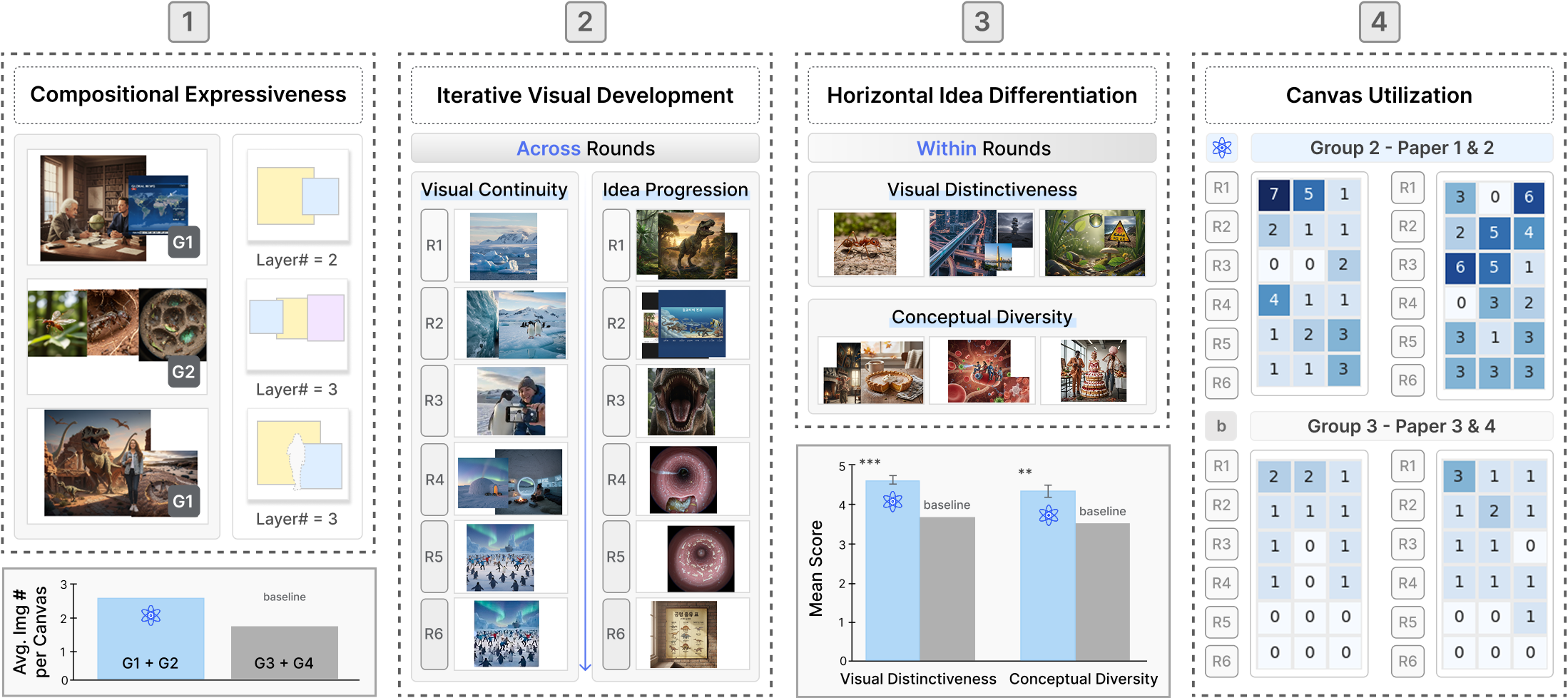}
  \caption{\textbf{Artifact characteristics of \atomix{}.} Examples from user study sessions illustrate key output differences. Left: \atomix{} produces canvases with higher (1) compositional expressiveness (more images per canvas) and greater (2) iterative visual development (higher visual continuity and idea progression scores). Right: \atomix{} yields greater (3) horizontal idea differentiation (higher visual distinctiveness and conceptual diversity scores) and more consistent (4) canvas utilization (fewer empty canvases).}
  \label{fig:artifact}
\end{figure*}

\subsubsection{Compositional Expressiveness}
\label{sec:compositional}

\begin{figure}
  \centering
  \includegraphics[width=1\linewidth]{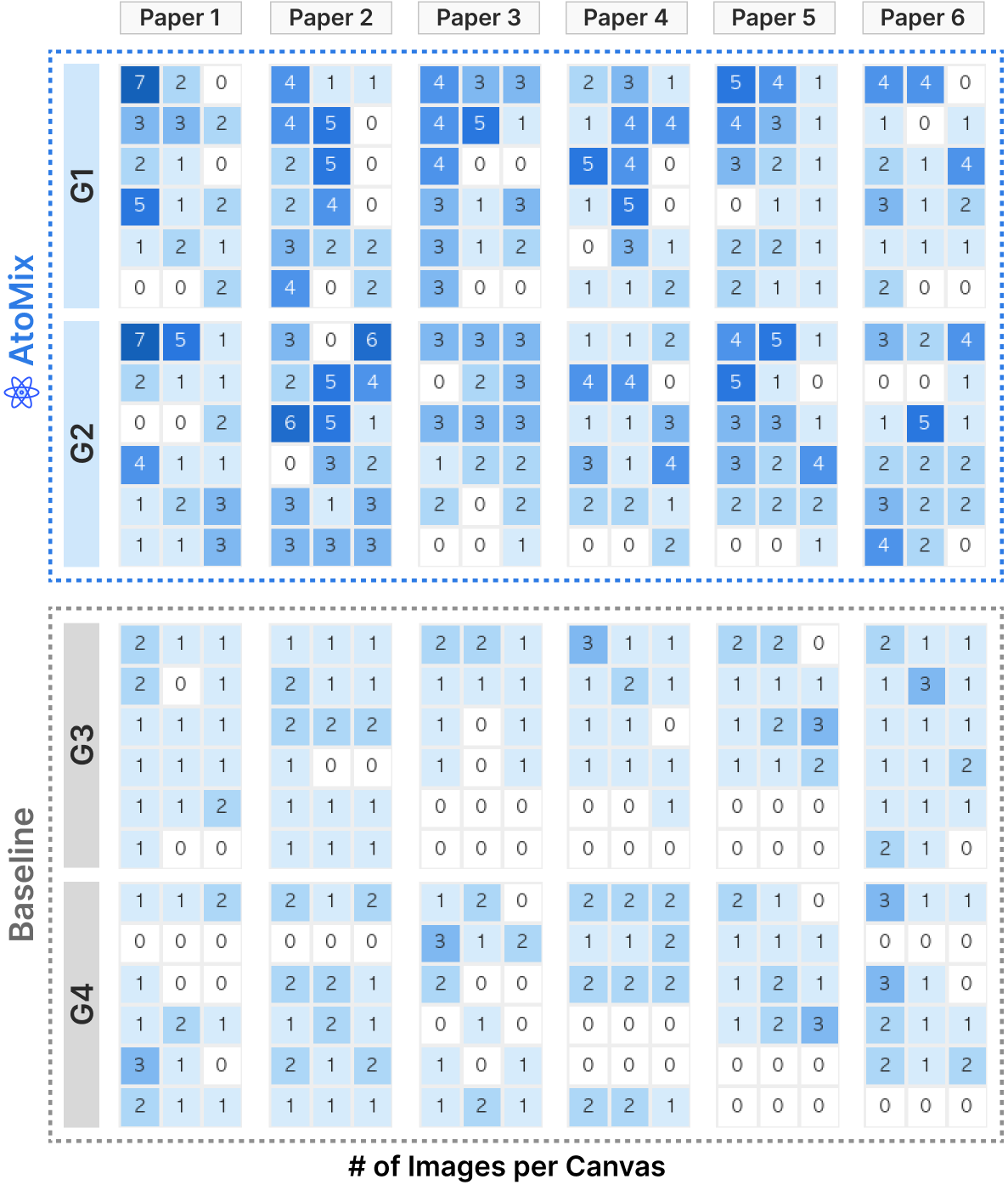}
  \caption{\textbf{Heatmap of image counts per canvas across all groups.} Each square represents a single canvas; there are 18 canvases per paper and 6 papers per brainwriting session, yielding 108 canvases per group in total. Darker colors indicate canvases with more images. \atomix{} exhibits consistently higher image counts across participants and rounds compared to the baseline.}
  \label{fig:heatmap}
\end{figure}

We quantified compositional expressiveness as the number of images added to each canvas, computed from interaction logs across all rounds. Participants using \atomix{} added significantly more images per canvas (M=9.5, SD=3.26) than baseline participants (M=4.42, SD=3.06, p<.001***) across the six rounds. This difference was consistent across all six rounds. Figure~\ref{fig:heatmap} shows the distribution of image counts per canvas, with higher densities in the \atomix{} condition than in the baseline.

Post-study interviews show that participants in the \atomix{} condition frequently placed multiple generated images on a single canvas. Six participants described workflows involving generating, layering, and arranging elements to iteratively build a concept: “I generated new images with the prompt and then combined them” (P2). Three participants used a background-first strategy, adding details incrementally (P1). In contrast, baseline participants engaged in limited collaging, often defaulting to a single image because they did not consider combining multiple images (P21, P22). While some participants added multiple images, these were typically placed side by side rather than composed into a unified visual: “I added additional images with similar concepts,” but “didn’t compose one big image from two” (P20).

\subsubsection{Iterative Visual Development}
\label{sec:iterative}

\begin{figure} 
  \centering
  \includegraphics[width=0.5\textwidth]{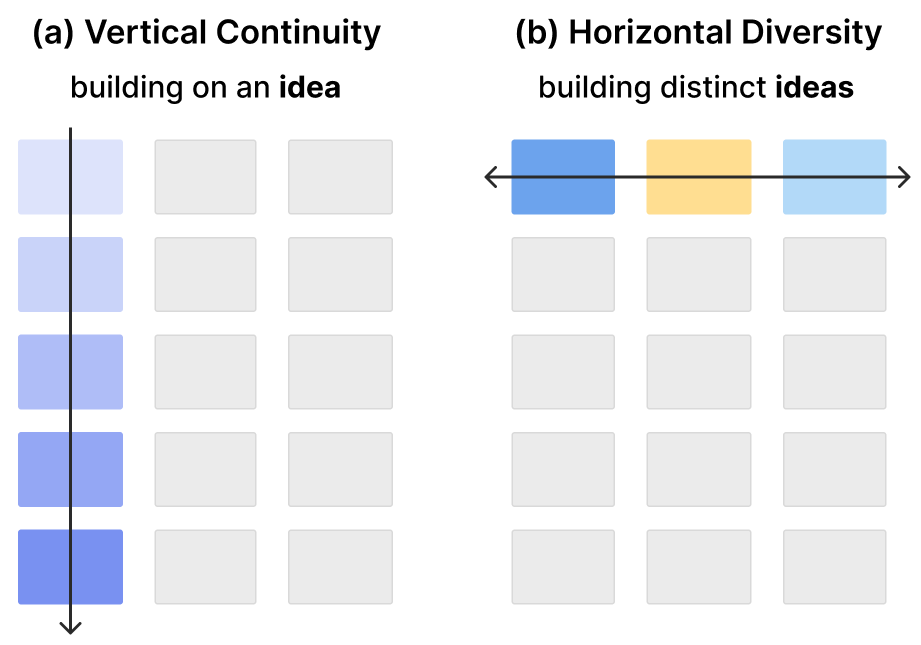}
  \caption{\textbf{Iterative visual development and horizontal idea differentiation.} (a) Vertical continuity captures how ideas evolve across rounds within a single column—measuring visual continuity and idea progression as successive participants build on prior work. (b) Horizontal differentiation captures how ideas vary across parallel canvases within the same round—measuring visual distinctiveness and conceptual diversity among concurrent contributions.}
  \label{fig:vertical_horizontal}
\end{figure}

We examined iterative visual development by assessing how ideas evolved across rounds within each column (Figure~\ref{fig:vertical_horizontal}a). LLM evaluations indicated higher visual continuity (M=3.72 vs. 3.44) and idea progression (M=3.56 vs. 3.19) compared to the \baseline{}; however, these differences were modest and did not reach statistical significance (Table~\ref{tab:vertical-horizontal-llm}), so they should be interpreted as directional trends rather than confirmed effects. Examples of iterative visual development patterns are shown in Figure~\ref{fig:artifact} (2).

Post-study interviews indicated that participants using \atomix{} actively built on prior ideas by scanning existing work and adding elements intended for later integration. As P2 noted, “I just added similar-feeling elements and thought maybe the next person would combine them.” This often produced a relay- or story-like flow, with each participant extending the previous contribution (P7). While some felt pressure to converge (P7), others experienced this shared narrative as collaborative momentum---“doing it together because there was a story” (P10). In contrast, iteration in the baseline condition was less structured and cumulative. Early rounds were highly divergent, with participants generating ideas independently rather than building on prior work (P17). Although some later referenced earlier ideas (P14), many struggled to sustain coherence, with core ideas becoming “diluted over time” (P14, P15, P18).

\begin{table*}[t] 
    \centering
    \small
    \begin{tabular}{lcccc}
    \toprule
    \multicolumn{5}{l}{\textbf{Vertical: Iterative Visual Development}} \\
    \midrule
    Metric & \atomix{} M (SD) & \baseline{} M (SD) & p-value & Cohen's d \\
    \midrule
    Visual Continuity    & 3.72 (1.23) & 3.44 (1.03) & 0.1467 (n.s.) & 0.245 \\
    Idea Progression     & 3.56 (1.11) & 3.19 (1.17) & 0.1678 (n.s.) & 0.318 \\
    \midrule
    \multicolumn{5}{l}{\textbf{Horizontal: Idea Differentiation}} \\
    \midrule
    Visual Distinctiveness & 4.61 (0.86) & 3.67 (1.71) & 0.0007*** & 0.696 \\
    Conceptual Diversity   & 4.35 (0.98) & 3.53 (1.64) & 0.0037**  & 0.608 \\
    \bottomrule
    \end{tabular}
    \vspace{6pt}
    \caption{LLM evaluation of vertical (iterative development) and horizontal (idea differentiation) metrics: Mann–Whitney U test results comparing \atomix{} and \baseline{}.}
    \label{tab:vertical-horizontal-llm}
\end{table*}

\subsubsection{ Horizontal Idea Differentiation}
\label{sec:horizontal}

We assessed horizontal idea differentiation by evaluating visual distinctiveness and conceptual diversity across parallel idea slots within each row (Figure~\ref{fig:vertical_horizontal}b). LLM-based evaluations showed higher visual distinctiveness (M = 4.61 vs. 3.67, p < .001***) and conceptual diversity (M = 4.35 vs. 3.53, p < .01**), leading to more divergent ideas in the \atomix{} condition compared to the baseline (Table~\ref{tab:vertical-horizontal-llm}). Standard deviations were lower for \atomix{} than the baseline for both metrics (distinctiveness: SD = 0.86 vs. 1.71; diversity: SD = 0.98 vs. 1.64).

Interview data indicate that participants in the \atomix{} condition sustained divergence by generating multiple ideas in parallel within each round. Some reported distributing effort across canvases rather than fully developing a single idea (i.e., “I couldn't fully develop all three ideas with good quality,” P2). Participants also described comparing ideas across canvases (P7) and using the cross-pollination feature to introduce variation in their ideas, as P9 noted, the “lightbulb (image-mashup) feature randomly selected canvases and gave really quirky new ideas.”

\subsubsection{Canvas Utilization}
\label{sec:canvas}

To examine canvas utilization, we analyzed interaction logs capturing empty canvases left unmodified at the end of each round. The log data show that \atomix{} participants left significantly fewer empty canvases (37 out of 216, 17.1\%) than \baseline{} participants (62 out of 216, 28.7\%) across the six rounds (p<.01**).

Post-study interviews indicate that participants in the \atomix{} often completed canvases under time pressure by copying, reusing, and extending existing images rather than leaving canvases blank. As one participant noted, “I just copied existing elements and added things on top” (P1), while others described continuing the existing flow through cross-pollination (P5, P8). In contrast, \baseline{} participants frequently left blank or partially completed canvases, often due to time constraints, image generation delays, or workflow fragmentation. Participants described “moving on without finishing images” (P14, P18) due to difficulty generating visuals that matched their intent (P23) or losing time while switching tools (P18).

\subsection{RQ2. How do interaction patterns in \atomix{} and the \baseline{} system influence ideation artifacts?}
\begin{figure*}[t]
  \centering
    \includegraphics[width=\textwidth]
    {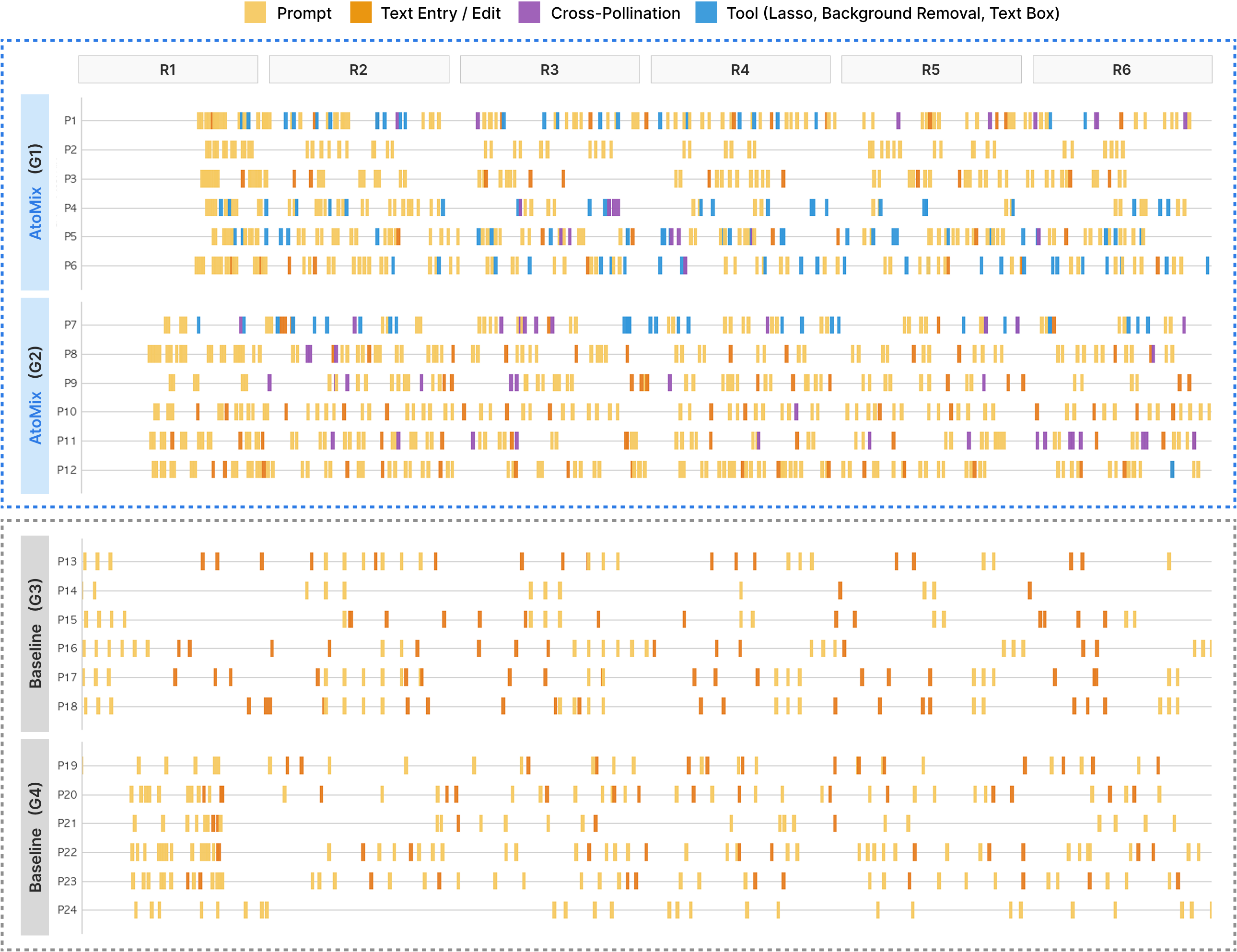}
  \caption{\textbf{User action sequences} across rounds in \atomix{} and \baseline{} conditions. Each row represents a participant's sequence of actions over six rounds, with colors indicating different action types (i.e., prompt submission, text entry, and tool usage). \atomix{} participants exhibit more frequent and varied actions, reflecting higher engagement and interaction complexity compared to \baseline{} participants.}
  \label{fig:sequence}
\end{figure*}

Having characterized the outputs, we next examine interaction practices associated with each system. Using interaction logs and post-study interviews, we analyze how participants generated, modified, combined, and communicated ideas across rounds. This analysis identifies three recurring interaction patterns associated with differences in system affordances. Figure~\ref{fig:sequence} shows the full action sequences for each participant across rounds.

\subsubsection{Atomic Prompting}
\label{sec:atomic}

\begin{figure}
  \centering
  \includegraphics[width=1\linewidth]{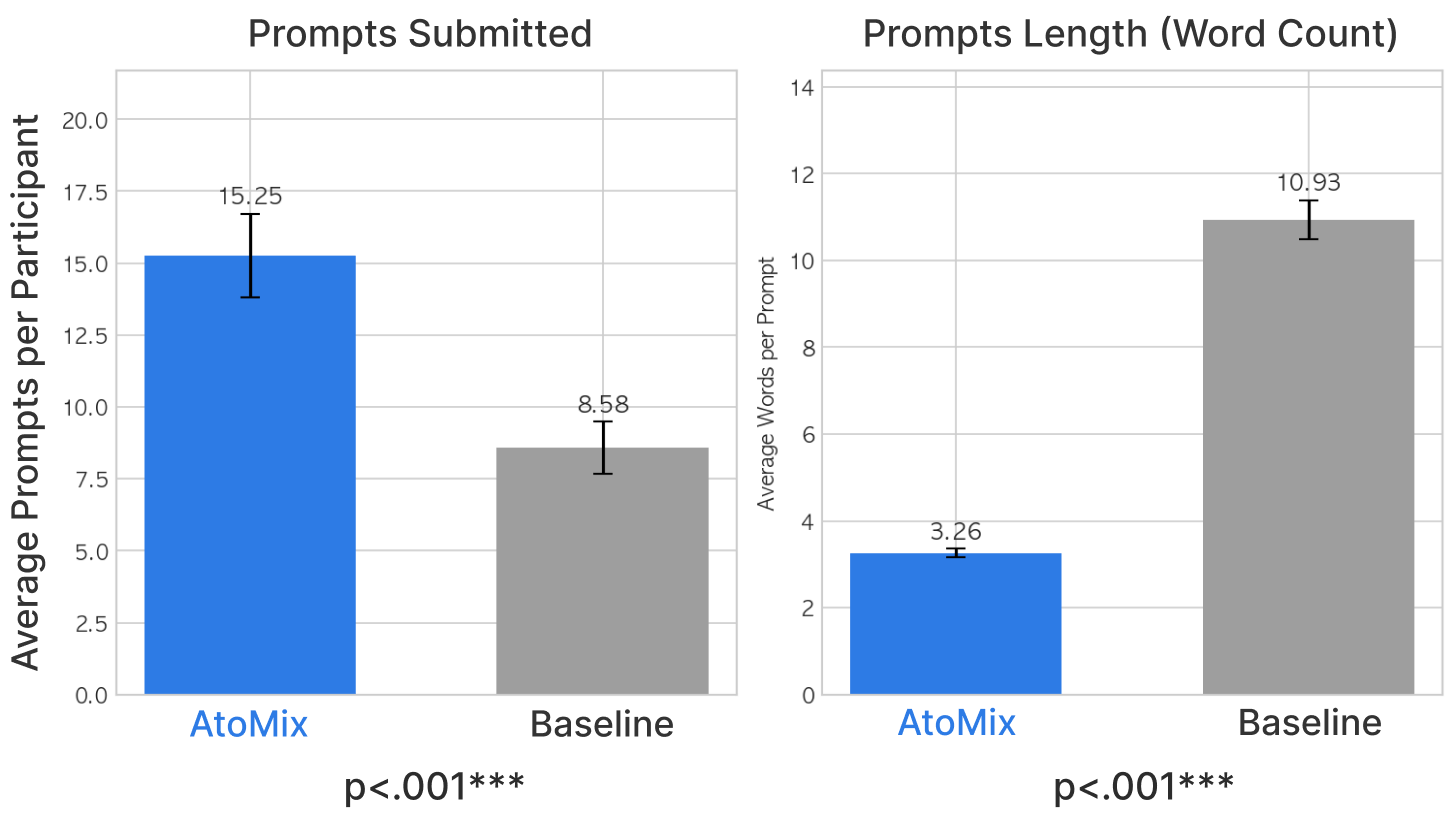}
  \caption{\textbf{Prompting behavior comparison.} (Left) Average Prompts per Participant. (Right) Average Word Count per Prompt. \atomix{} users submitted significantly more prompts with shorter average lengths compared to \baseline{} users.}
  \label{fig:prompt}
\end{figure}

We examined prompting behavior using interaction logs capturing prompt counts and lengths across all rounds. The log data show that participants using \atomix{} submitted significantly more prompts (M=15.25, SD=5.03) than \baseline{} participants (M=8.58, SD=3.14) across the six rounds (p<.001***). Additionally, prompts in the \atomix{} condition were shorter on average (M=3.26 words, SD=1.94) compared to \baseline{} (M=10.93 words, SD=7.61) (p<.001***). Figure~\ref{fig:prompt} illustrates these differences in prompting behavior.

Post-study interview data indicate that participants in the ~\textsc{Ato\-Mix} condition frequently submitted short prompts focused on individual elements as they found detailed prompts ineffective, as P1 noted, “If I input a very long description… it doesn't work or takes too long, so I just input the background only.” This granular approach was especially common early in ideation, when participants “just made things as they came” (P11), enabling rapid exploration. In contrast, \baseline{} participants followed an iterative prompt-refinement pattern, repeatedly editing and reusing a single prompt. Participants described starting with an initial description and incrementally modifying it based on generated outputs (P15); as P16 explained, “I would just take the previous description as my prompt, edit it, generate the image, and bring it over.”

\subsubsection{Granular Editing}
\label{sec:granular}

We examined granular editing behavior using interaction logs capturing tool use and image-guided generation. \atomix{} supports granular editing through tool-based operations (e.g., lasso selection and background removal) and image-guided generation combining text prompts with existing images. Log data show that \atomix{} participants used tool-based operations 140 times (104 background removals, 23 lasso selections, and 13 text box uses). In addition, 46 prompts (12\% of all \atomix{} prompts) used image-guided generation, combining visual references with text instructions.

Post-study interviews indicated that granular editing enabled participants to iteratively build on prior images by targeting specific elements for change. For example, P7 explained, “If I had an idea I could add to the previous one, I would screenshot (lasso) it and say something like ‘modify this part.’” In contrast, using the \baseline{} system, participants frequently expressed a desire for granular editing. They noted that being unable to edit existing images or work on the same canvas limited their ability to express and refine ideas: “If I could create on the same page or edit existing images, I think I could express my thoughts better” (P18). Participants reported that this approach affected visual consistency, as “regenerating images from scratch made style and overall image coherence suffer when modifications needed were small” (P23). Reflecting on this gap, they emphasized the need for fine-tuning capabilities once ideas were established, such as “fine-tuning options to adjust detailed image elements” or modifying “only specific parts” (P21).

\subsubsection{Cross-Pollination Behaviors}
\label{sec:cross-pollination}

We examined cross-pollination behaviors using interaction logs capturing image reuse, references to prior canvases, and cross-canvas operations across rounds. Log data show that participants performed 58 cross-pollination actions in the \atomix{} condition, including use of the image mashup feature and object paste actions from other canvases.

Interview data indicate that participants in the \atomix{} condition reused and extended content from prior canvases. They treated prior work as references, making incremental additions rather than starting from scratch. Participants described using combination and idea-generation features to mix ideas across boards; for example, “I mixed things from other boards” (P4). Others extended ideas through lightweight reworking, such as copy-pasting an image and adding small elements (P8). In contrast, \baseline{} participants built on others’ ideas less directly and with greater effort. They expressed a need for better support to reuse and extend prior work; as P17 noted, “It would be much better if I could utilize previous people's ideas or generated images to supplement my own ideas.” Time pressure further limited cross-pollination, with participants feeling compelled to create new content rather than reuse prior work  (P15).



\subsection{RQ3. How do participants perceive and experience \atomix{} for  visual brainwriting?}

To address RQ3, we analyzed participants’ experiences with \atomix{} during collaborative visual brainwriting using post-study interviews. Our analysis identified four key themes, which capture how \atomix{} shaped both collaborative dynamics and participants’ subjective experience of ideation.

\subsubsection{Workflow Integration}
\label{sec:workflow}

Our thematic analysis revealed that participants valued \atomix{}’s integrated workflow, which allowed them to search, generate, and combine images within a single environment. Participants appreciated being able to work without switching tools, describing this as “faster and less hectic than relying on external apps” (P2, P7), which supported the rapid generation of multiple ideas (P5). Participants also noted the usefulness of contextual features, such as “viewing the prompts behind generated images to aid interpretation” (P5). Contrarily, in the \baseline{} condition, participants described a fragmented and cumbersome workflow that disrupted ideation flow. Moving between writing ideas and generating images was seen as inefficient; “writing ideas then having to move to generate images felt a bit cumbersome” (P18). Participants expressed a desire for tighter integration, suggesting that being able to generate images directly alongside text “on one page” would improve immersion and efficiency (P18). They highlighted friction in reusing prior work, such as having to retype prompts (P21) or being unable to easily build on others’ images 

\subsubsection{Collaborative Experience}
\label{sec:collaborative}

Our results revealed that participants using \atomix{}  frequently described the process as collaborative, cumulative, and story-like (P1, P3, P7, P10, P12). Participants enjoyed seeing ideas build over time, noting that “everything accumulated” and “became more detailed” (P1), and often felt they were responding directly to others’ contributions which made collaboration feel like “passing a ball back and forth” (P1). In contrast, \baseline{} participants more often experienced the task as individual work performed in sequence rather than true co-creation. While some acknowledged moments of collaboration---"especially early divergence” (P19) or when “referencing others’ images” (P20) ---many reflected that they were “generating ideas individually rather than developing ideas together” (P18), or “continuously performing given tasks” (P16). Differences in personal priorities led to diluted core ideas (P14, P13), and time pressure further reduced awareness of others’ presence (P20). 

\subsubsection{Engagement Through Immersion and Flow}
\label{sec:engagement}

Participants noted that \atomix{} generally fostered stronger engagement, with frequently describing the experience as “fun” (P3) and “immersive” (P11, P9). The integrated workflow helped maintain focus, as P3 noted that “search for images and combine ideas within one window supported deeper thinking”. Time pressure often enhanced flow, making the process feel “more like a game” and encouraging rapid ideation (P11, P10). In the \baseline{} condition, participants also experienced immersion, particularly due to time constraints, as P20 noted that “having to create three ideas within the 5-minute time constraint really increased immersion”. However, engagement was reported to be more fragile and frequently interrupted by workflow friction, such as “going back and forth between image generation and canvas” (P16), retyping prompts, or waiting for images to load (P13). Participants noted that these interruptions made it harder to maintain flow and regain focus once broken (P18). 

\subsubsection{AI Control and Creative Agency}
\label{sec:ai-control}

Our analysis revealed that \atomix{}'s image quality was sufficient for early ideation but offered limited control. Participants found it effective for rough idea generation (P5, P6) but noted that outputs were at times uniform or “deviated from intent” (P9), making ideas harder to concretize (P10, P12). This lack of control sometimes reduced the creative agency. In the \baseline{} condition, participants also used AI to quickly generate initial visuals, but faced similar control limitations. AI outputs often captured only “about 80\% of intent (P14)”, lacked text support (P16), and were rarely refined further once acceptable (P19).

\paragraph{\textbf{Summary of Findings}}
Across three research questions, our results show that \atomix{} produced more compositionally rich canvases with greater horizontal differentiation and fewer empty slots, while vertical continuity showed a positive but non-significant trend. These artifact differences were associated with three interaction patterns: atomic prompting, granular editing, and cross-pollination, enabled by \atomix{}'s fragment-level affordances. Participants valued the integrated workflow and collaborative momentum, though both conditions faced similar limitations in AI controllability.

%% file: sections/discussion.tex
\subsection{From Outputs to Materials}

Our artifact analysis shows that canvases created in the \atomix{} condition differed from those in the baseline along three key dimensions: they featured more compositionally rich collages, showed higher variety of ideas across parallel canvases and increased canvas utilization. We interpret these differences as consequences of reframing generative AI outputs from \emph{finished products} to \emph{manipulable materials}.

The increase in multi-element collages appears linked to \atomix{}'s affordances—drag-and-drop placement, editing tools, and layering—which reduced the commitment required per image~(\S\ref{sec:compositional}). In the baseline, each generation was a high-stakes attempt to capture a full concept; partial matches felt like wasted effort. \atomix{} reframed partial fragments as useful building blocks, allowing participants to generate a background, then a foreground object, then an annotation—assembling meaning incrementally rather than demanding it from a single prompt. This aligns with Tversky's observation that external representations become thinking tools when manipulated piecemeal~\cite{tversky_visualizing_2011}. The same low-commitment workflow explains vertical continuity~(\S\ref{sec:iterative}): when prior contributions can be selectively preserved or recombined rather than regenerated wholesale, participants have structural incentives to build on existing work. Horizontal differentiation~(\S\ref{sec:horizontal}) likewise increased because branching into a new direction cost only a short prompt and a drag operation, leading to more divergent ideas. Even under time pressure, participants could copy, extend, or remix existing fragments rather than leave canvases blank, preserving ideation momentum and reducing the empty canvases that interrupted baseline sessions~(\S\ref{sec:canvas}).

Across all four dimensions, the common thread is treating generated images as materials that are partial, combinable, and low-cost to modify. This unlocked cumulative creative processes that monolithic generation impedes. That said, \atomix{}'s integrated workflow also reduces context switching relative to the baseline, so we cannot rule out that some of the observed benefits may stem from integration itself rather than the compositional framing alone.

\subsection{Atomicity, Granularity, and Cross-pollination}

The interaction patterns help explain \emph{how} the observed artifact differences emerged. Participants learned that short, single-element prompts were more reliable and faster than long, scene-level descriptions; rather than attempting to craft a single optimal prompt, they issued many small requests and assembled the results spatially~(\S\ref{sec:atomic}). This behavior parallels least-to-most prompting strategies in language models~\cite{zhou2023leasttomostpromptingenablescomplex} and cognitive chunking in creative problem-solving~\cite{finke_creative_1992}: decomposing a complex goal into controllable subgoals reduces cognitive load while keeping alternatives open longer. In contrast, the baseline’s prompt-refinement loop---edit, regenerate, evaluate, repeat---tended to lock participants into a single conceptual trajectory, whereas atomic prompting supported parallel exploration.

These atomic prompts fed into granular editing, enabling in-place transformation rather than global regeneration~(\S\ref{sec:granular}). By lassoing a region and requesting a local change, participants preserved surrounding context and visual coherence, operationalizing Schön's "seeing--moving--seeing" cycle~\cite{schon1992reflective}: seeing a partial result, making a situated adjustment, and immediately observing its consequences. In the baseline, even minor revisions required re-describing entire scenes, disrupting continuity and raising the cost of iteration. Over time, cross-pollination increased as a growing critical mass of fragments made recomposition more attractive than generation from scratch~(\S\ref{sec:cross-pollination}). \atomix{}’s mashup feature and copy–paste affordances lowered the friction of such recombination, turning the shared canvas into a reservoir of reusable parts.

Together, these interaction patterns suggest a design principle: \emph{reduce the cost of small, reversible moves}. When generation, modification, and recombination are cheap, participants explore more widely, iterate more frequently, and build on each other's work more readily, producing the artifact characteristics observed in RQ1.

\subsection{Implications for GenAI-Augmented Ideation}

The challenges that GenAI-augmented ideation often faces—intent drift requiring repeated regeneration~\cite{Peter2025,Jiaxin_2025} and early convergence toward generic outputs~\cite{holzner2025generativeaicreativitysystematic,Joongi2025}—may stem less from GenAI itself than from how it is typically integrated. Standard workflows frame each generation as a complete deliverable to accept, reject, or refine—a monolithic interaction that encourages anchoring and amplifies convergence across participants. \atomix{}'s compositional approach can help mitigate this pattern: by treating generated images as partial materials to be decomposed, combined, and extended, participants introduced variation at the \emph{assembly} stage rather than relying solely on prompt-level divergence. This produced greater vertical continuity alongside horizontal differentiation, suggesting that cumulative build-on and sustained divergence can coexist under low-cost recombination.

This reframing extends prior work on AI-augmented brainwriting, which showed improvements in fluency, elaboration, and divergence–convergence scaffolding in text-based ideation~\cite{shaer_ai-augmented_2024}. Our findings indicate that similar benefits can be realized in visual brainwriting, but only when AI outputs are treated not as completed images but as manipulable components. More broadly, these findings suggest that sustaining diversity and supporting user intent in GenAI creativity tools may depend less on model capability alone than on interaction design—specifically, on whether interfaces frame AI outputs as endpoints or as starting points for further transformation. 

We note that because both conditions used the 6-3-5 structure, some effects may reflect tool–method interactions—for example, timed rotations may amplify the benefit of low-cost fragment reuse more than they would in a free-form setting. Still, returning to brainwriting's original promise—parallel idea generation, silent exchange, and cumulative build-on~\cite{heslin_better_2009}—our study suggests that GenAI can amplify rather than undermine these strengths when integrated at the right level of granularity. The 6-3-5 structure's timed rotations and shared sheets were designed to encourage reinterpretation and incremental extension; \atomix{}'s fragment-based workflow aligns with these principles by making prior contributions visually accessible and easily remixable. Rather than replacing human ideation with GenAI, compositional workflows position AI as a material supplier that accelerates externalization while preserving the collaborative, cumulative dynamics that make visual brainwriting effective. Building on this perspective, future work could personalize compositional workflows by passively identifying individual contributors during sessions \cite{saad_sensing_2023}, enabling systems to adapt recombination suggestions and cross-pollination opportunities to each participant's contribution history and ideation style.

%% file: sections/conclusion.tex
We present \atomix{}, a GenAI-augmented collaborative visual brainwriting system for remote users that treats generated images as compositional building blocks rather than complete deliverables. This design supports atomic prompting, granular editing, and cross-pollination, addressing bottlenecks identified in our formative study where prompt–regenerate cycles disrupted incremental, collaborative ideation. A comparative study with 24 participants showed that \atomix{} enabled the creation of more compositionally expressive canvases with greater idea differentiation, while still supporting smooth, incremental progression. Participants shifted from crafting long prompts to issuing short, iterative atomic prompts that assembled ideas from visual fragments. Region-based editing and cross-canvas mashups further encouraged reuse and co-creation. Together, these findings demonstrate that GenAI can better support creative practice by shifting from monolithic image generation toward compositional assembly---sustaining cumulative idea development in collaborative design ideation.

%% file: main.bbl

\begin{thebibliography}{63}


\ifx \showCODEN    \undefined \def \showCODEN     #1{\unskip}     \fi
\ifx \showISBNx    \undefined \def \showISBNx     #1{\unskip}     \fi
\ifx \showISBNxiii \undefined \def \showISBNxiii  #1{\unskip}     \fi
\ifx \showISSN     \undefined \def \showISSN      #1{\unskip}     \fi
\ifx \showLCCN     \undefined \def \showLCCN      #1{\unskip}     \fi
\ifx \shownote     \undefined \def \shownote      #1{#1}          \fi
\ifx \showarticletitle \undefined \def \showarticletitle #1{#1}   \fi
\ifx \showURL      \undefined \def \showURL       {\relax}        \fi
\providecommand\bibfield[2]{#2}
\providecommand\bibinfo[2]{#2}
\providecommand\natexlab[1]{#1}
\providecommand\showeprint[2][]{arXiv:#2}

\bibitem[Bae et~al\mbox{.}(2020)]%
        {Bae2020Spinneret}
\bibfield{author}{\bibinfo{person}{Suyun Bae}, \bibinfo{person}{Oh-Hyun Kwon}, \bibinfo{person}{Senthil Chandrasegaran}, {and} \bibinfo{person}{Kwan-Liu Ma}.} \bibinfo{year}{2020}\natexlab{}.
\newblock \showarticletitle{Spinneret: Aiding Creative Ideation through Non-Obvious Concept Associations}. In \bibinfo{booktitle}{\emph{Proceedings of the CHI Conference on Human Factors in Computing Systems}}.
\newblock
\href{https://doi.org/10.1145/3313831.3376746}{doi:\nolinkurl{10.1145/3313831.3376746}}
\newblock
\shownote{CHI '20}.


\bibitem[Boden(2004)]%
        {boden_creative_2004}
\bibfield{author}{\bibinfo{person}{Margaret~A. Boden}.} \bibinfo{year}{2004}\natexlab{}.
\newblock \bibinfo{booktitle}{\emph{The Creative Mind: Myths and Mechanisms} (\bibinfo{edition}{2nd} ed.)}.
\newblock \bibinfo{publisher}{Routledge}, \bibinfo{address}{London}.
\newblock


\bibitem[Bouschery et~al\mbox{.}(2023)]%
        {bouschery_ai-augmented_2023}
\bibfield{author}{\bibinfo{person}{Sebastian~Gregor Bouschery}, \bibinfo{person}{Vera Blazevic}, {and} \bibinfo{person}{Frank~T. Piller}.} \bibinfo{year}{2023}\natexlab{}.
\newblock \showarticletitle{{AI}-{Augmented} {Creativity}: {Overcoming} the {Productivity} {Loss} in {Brainstorming} {Groups}}.
\newblock \bibinfo{journal}{\emph{Academy of Management Proceedings}} \bibinfo{volume}{2023}, \bibinfo{number}{1} (\bibinfo{date}{Aug.} \bibinfo{year}{2023}), \bibinfo{pages}{11938}.
\newblock
\showISSN{0065-0668}
\href{https://doi.org/10.5465/AMPROC.2023.11938abstract}{doi:\nolinkurl{10.5465/AMPROC.2023.11938abstract}}
\newblock
\shownote{Publisher: Academy of Management}.


\bibitem[Braun and Clarke(2006)]%
        {braun_using_2006}
\bibfield{author}{\bibinfo{person}{Virginia Braun} {and} \bibinfo{person}{Victoria Clarke}.} \bibinfo{year}{2006}\natexlab{}.
\newblock \showarticletitle{Using thematic analysis in psychology}.
\newblock \bibinfo{journal}{\emph{Qualitative Research in Psychology}} \bibinfo{volume}{3}, \bibinfo{number}{2} (\bibinfo{year}{2006}), \bibinfo{pages}{77--101}.
\newblock
\href{https://doi.org/10.1191/1478088706qp063oa}{doi:\nolinkurl{10.1191/1478088706qp063oa}}


\bibitem[Cai et~al\mbox{.}(2023)]%
        {10.1145/3582269.3615596}
\bibfield{author}{\bibinfo{person}{Alice Cai}, \bibinfo{person}{Steven~R Rick}, \bibinfo{person}{Jennifer~L Heyman}, \bibinfo{person}{Yanxia Zhang}, \bibinfo{person}{Alexandre Filipowicz}, \bibinfo{person}{Matthew Hong}, \bibinfo{person}{Matt Klenk}, {and} \bibinfo{person}{Thomas Malone}.} \bibinfo{year}{2023}\natexlab{}.
\newblock \showarticletitle{DesignAID: Using Generative AI and Semantic Diversity for Design Inspiration}. In \bibinfo{booktitle}{\emph{Proceedings of The ACM Collective Intelligence Conference}} (Delft, Netherlands) \emph{(\bibinfo{series}{CI '23})}. \bibinfo{publisher}{Association for Computing Machinery}, \bibinfo{address}{New York, NY, USA}, \bibinfo{pages}{1–11}.
\newblock
\showISBNx{9798400701139}
\href{https://doi.org/10.1145/3582269.3615596}{doi:\nolinkurl{10.1145/3582269.3615596}}


\bibitem[Cao et~al\mbox{.}(2015)]%
        {cao_idea_2015}
\bibfield{author}{\bibinfo{person}{Jill Cao}, \bibinfo{person}{Scott~D. Fleming}, \bibinfo{person}{Margaret Burnett}, {and} \bibinfo{person}{Christopher Scaffidi}.} \bibinfo{year}{2015}\natexlab{}.
\newblock \showarticletitle{Idea {Garden}: {Situated} {Support} for {Problem} {Solving} by {End}-{User} {Programmers}}.
\newblock \bibinfo{journal}{\emph{Interacting with Computers}} \bibinfo{volume}{27}, \bibinfo{number}{6} (\bibinfo{date}{Nov.} \bibinfo{year}{2015}), \bibinfo{pages}{640--660}.
\newblock
\showISSN{0953-5438, 1873-7951}
\href{https://doi.org/10.1093/iwc/iwu022}{doi:\nolinkurl{10.1093/iwc/iwu022}}


\bibitem[Choi et~al\mbox{.}(2024)]%
        {choi_creativeconnect_2024}
\bibfield{author}{\bibinfo{person}{DaEun Choi}, \bibinfo{person}{Sumin Hong}, \bibinfo{person}{Jeongeon Park}, \bibinfo{person}{John Joon~Young Chung}, {and} \bibinfo{person}{Juho Kim}.} \bibinfo{year}{2024}\natexlab{}.
\newblock \showarticletitle{{CreativeConnect}: {Supporting} {Reference} {Recombination} for {Graphic} {Design} {Ideation} with {Generative} {AI}}. In \bibinfo{booktitle}{\emph{Proceedings of the 2024 {CHI} {Conference} on {Human} {Factors} in {Computing} {Systems}}} \emph{(\bibinfo{series}{{CHI} '24})}. \bibinfo{publisher}{Association for Computing Machinery}, \bibinfo{address}{New York, NY, USA}, \bibinfo{pages}{1--25}.
\newblock
\showISBNx{979-8-4007-0330-0}
\href{https://doi.org/10.1145/3613904.3642794}{doi:\nolinkurl{10.1145/3613904.3642794}}


\bibitem[Dalsgaard(2025)]%
        {Peter2025}
\bibfield{author}{\bibinfo{person}{Peter Dalsgaard}.} \bibinfo{year}{2025}\natexlab{}.
\newblock \showarticletitle{Creative Ambiguity and Cognitive Tension in Generative AI Tools}. In \bibinfo{booktitle}{\emph{Proceedings of the 36th Annual Conference of the European Association of Cognitive Ergonomics}} \emph{(\bibinfo{series}{ECCE '25})}. \bibinfo{publisher}{Association for Computing Machinery}, \bibinfo{address}{New York, NY, USA}, Article \bibinfo{articleno}{4}, \bibinfo{numpages}{10}~pages.
\newblock
\showISBNx{9798400720338}
\href{https://doi.org/10.1145/3746175.3746183}{doi:\nolinkurl{10.1145/3746175.3746183}}


\bibitem[Finke et~al\mbox{.}(1992)]%
        {finke_creative_1992}
\bibfield{author}{\bibinfo{person}{Ronald~A. Finke}, \bibinfo{person}{Thomas~B. Ward}, {and} \bibinfo{person}{Steven~M. Smith}.} \bibinfo{year}{1992}\natexlab{}.
\newblock \bibinfo{booktitle}{\emph{Creative {Cognition}: {Theory}, {Research}, and {Applications}}}.
\newblock \bibinfo{publisher}{The MIT Press}.
\newblock
\showISBNx{978-0-262-27265-0}
\href{https://doi.org/10.7551/mitpress/7722.001.0001}{doi:\nolinkurl{10.7551/mitpress/7722.001.0001}}


\bibitem[Frich et~al\mbox{.}(2019)]%
        {frich_mapping_2019}
\bibfield{author}{\bibinfo{person}{Jonas Frich}, \bibinfo{person}{Lindsay MacDonald~Vermeulen}, \bibinfo{person}{Christian Remy}, \bibinfo{person}{Michael~Mose Biskjaer}, {and} \bibinfo{person}{Peter Dalsgaard}.} \bibinfo{year}{2019}\natexlab{}.
\newblock \showarticletitle{Mapping the {Landscape} of {Creativity} {Support} {Tools} in {HCI}}. In \bibinfo{booktitle}{\emph{Proceedings of the {CHI} {Conference} on {Human} {Factors} in {Computing} {Systems}}}.
\newblock
\href{https://doi.org/10.1145/3290605.3300619}{doi:\nolinkurl{10.1145/3290605.3300619}}


\bibitem[Geyer et~al\mbox{.}(2012)]%
        {geyer_ideavis_2012}
\bibfield{author}{\bibinfo{person}{Florian Geyer}, \bibinfo{person}{Jochen Budzinski}, {and} \bibinfo{person}{Harald Reiterer}.} \bibinfo{year}{2012}\natexlab{}.
\newblock \showarticletitle{{IdeaVis}: a hybrid workspace and interactive visualization for paper-based collaborative sketching sessions}. In \bibinfo{booktitle}{\emph{Proceedings of the 7th {Nordic} {Conference} on {Human}-{Computer} {Interaction}: {Making} {Sense} {Through} {Design}}}. \bibinfo{publisher}{ACM}, \bibinfo{address}{Copenhagen Denmark}, \bibinfo{pages}{331--340}.
\newblock
\showISBNx{978-1-4503-1482-4}
\href{https://doi.org/10.1145/2399016.2399069}{doi:\nolinkurl{10.1145/2399016.2399069}}


\bibitem[Guilford(1950)]%
        {guiford_creativity_1950}
\bibfield{author}{\bibinfo{person}{Joy~Paul Guilford}.} \bibinfo{year}{1950}\natexlab{}.
\newblock \showarticletitle{Creativity.}
\newblock \bibinfo{journal}{\emph{American psychologist}} \bibinfo{volume}{5}, \bibinfo{number}{9} (\bibinfo{year}{1950}), \bibinfo{pages}{444}.
\newblock
\href{https://doi.org/10.1037/h0063487}{doi:\nolinkurl{10.1037/h0063487}}


\bibitem[Haddad et~al\mbox{.}(2025)]%
        {haddad2025user}
\bibfield{author}{\bibinfo{person}{F.~G. Haddad}, \bibinfo{person}{K.~M. Jang}, \bibinfo{person}{F. Duarte}, {et~al\mbox{.}}} \bibinfo{year}{2025}\natexlab{}.
\newblock \showarticletitle{User-centered evaluation of visual generative AI for city design: An exploratory technology acceptance model analysis}.
\newblock \bibinfo{journal}{\emph{Cognition, Technology \& Work}} (\bibinfo{year}{2025}).
\newblock
\href{https://doi.org/10.1007/s10111-025-00816-7}{doi:\nolinkurl{10.1007/s10111-025-00816-7}}


\bibitem[Heslin(2009)]%
        {heslin_better_2009}
\bibfield{author}{\bibinfo{person}{Peter~A. Heslin}.} \bibinfo{year}{2009}\natexlab{}.
\newblock \showarticletitle{Better than brainstorming? {Potential} contextual boundary conditions to brainwriting for idea generation in organizations}.
\newblock \bibinfo{journal}{\emph{Journal of Occupational and Organizational Psychology}} \bibinfo{volume}{82}, \bibinfo{number}{1} (\bibinfo{year}{2009}), \bibinfo{pages}{129--145}.
\newblock
\showISSN{2044-8325}
\href{https://doi.org/10.1348/096317908X285642}{doi:\nolinkurl{10.1348/096317908X285642}}
\newblock
\shownote{\_eprint: https://onlinelibrary.wiley.com/doi/pdf/10.1348/096317908X285642}.


\bibitem[Holzner et~al\mbox{.}(2025)]%
        {holzner2025generativeaicreativitysystematic}
\bibfield{author}{\bibinfo{person}{Niklas Holzner}, \bibinfo{person}{Sebastian Maier}, {and} \bibinfo{person}{Stefan Feuerriegel}.} \bibinfo{year}{2025}\natexlab{}.
\newblock \bibinfo{title}{Generative AI and Creativity: A Systematic Literature Review and Meta-Analysis}.
\newblock
\showeprint[arxiv]{2505.17241}~[cs.HC]
\urldef\tempurl%
\url{https://arxiv.org/abs/2505.17241}
\showURL{%
\tempurl}


\bibitem[Huang et~al\mbox{.}(2007)]%
        {Huang2007IdeaStormingCube}
\bibfield{author}{\bibinfo{person}{Chun-Chieh Huang}, \bibinfo{person}{Tsai-Yen Li}, \bibinfo{person}{Hao-Chuan Wang}, {and} \bibinfo{person}{Chun-Yen Chang}.} \bibinfo{year}{2007}\natexlab{}.
\newblock \showarticletitle{A Collaborative Support Tool for Creativity Learning: Idea Storming Cube}. In \bibinfo{booktitle}{\emph{Proceedings of the 7th IEEE International Conference on Advanced Learning Technologies}}.
\newblock
\href{https://doi.org/10.1109/ICALT.2007.6}{doi:\nolinkurl{10.1109/ICALT.2007.6}}
\newblock
\shownote{ICALT '07}.


\bibitem[Jeon et~al\mbox{.}(2021)]%
        {jeon_fashionq_2021}
\bibfield{author}{\bibinfo{person}{Youngseung Jeon}, \bibinfo{person}{Seungwan Jin}, \bibinfo{person}{Patrick~C. Shih}, {and} \bibinfo{person}{Kyungsik Han}.} \bibinfo{year}{2021}\natexlab{}.
\newblock \showarticletitle{{FashionQ}: {An} {AI}-{Driven} {Creativity} {Support} {Tool} for {Facilitating} {Ideation} in {Fashion} {Design}}. In \bibinfo{booktitle}{\emph{Proceedings of the 2021 {CHI} {Conference} on {Human} {Factors} in {Computing} {Systems}}} \emph{(\bibinfo{series}{{CHI} '21})}. \bibinfo{publisher}{Association for Computing Machinery}, \bibinfo{address}{New York, NY, USA}, \bibinfo{pages}{1--18}.
\newblock
\showISBNx{978-1-4503-8096-6}
\href{https://doi.org/10.1145/3411764.3445093}{doi:\nolinkurl{10.1145/3411764.3445093}}


\bibitem[Jin et~al\mbox{.}(2021)]%
        {jin_creativesvg_2021}
\bibfield{author}{\bibinfo{person}{Yucheng Jin}, \bibinfo{person}{Ge Gao}, \bibinfo{person}{Tiffany Ong}, {and} \bibinfo{person}{Xi Su}.} \bibinfo{year}{2021}\natexlab{}.
\newblock \showarticletitle{{CreativeSVG}: {A} {Creativity} {Support} {Tool} for {Abstract} {Background} {Design} with {Generative} {Vector} {Graphics}}. In \bibinfo{booktitle}{\emph{Joint {Proceedings} of the {ACM} {IUI} 2021 {Workshops}}}. \bibinfo{pages}{1--6}.
\newblock
\urldef\tempurl%
\url{https://ceur-ws.org/Vol-2903/IUI21WS-HAIGEN-6.pdf}
\showURL{%
\tempurl}


\bibitem[Kang et~al\mbox{.}(2024)]%
        {kang_biospark_2024}
\bibfield{author}{\bibinfo{person}{Hyeonsu~B Kang}, \bibinfo{person}{David Chuan-En Lin}, \bibinfo{person}{Nikolas Martelaro}, \bibinfo{person}{Aniket Kittur}, \bibinfo{person}{Yan-Ying Chen}, {and} \bibinfo{person}{Matthew~K. Hong}.} \bibinfo{year}{2024}\natexlab{}.
\newblock \showarticletitle{{BioSpark}: {An} {End}-to-{End} {Generative} {System} for {Biological}-{Analogical} {Inspirations} and {Ideation}}. In \bibinfo{booktitle}{\emph{Extended {Abstracts} of the {CHI} {Conference} on {Human} {Factors} in {Computing} {Systems}}} \emph{(\bibinfo{series}{{CHI} {EA} '24})}. \bibinfo{publisher}{Association for Computing Machinery}, \bibinfo{address}{New York, NY, USA}, \bibinfo{pages}{1--13}.
\newblock
\showISBNx{979-8-4007-0331-7}
\href{https://doi.org/10.1145/3613905.3651035}{doi:\nolinkurl{10.1145/3613905.3651035}}


\bibitem[Kazi et~al\mbox{.}(2011)]%
        {kazi_sandcanvas_2011}
\bibfield{author}{\bibinfo{person}{Rubaiat~Habib Kazi}, \bibinfo{person}{Kien~Chuan Chua}, \bibinfo{person}{Shengdong Zhao}, \bibinfo{person}{Richard Davis}, {and} \bibinfo{person}{Kok-Lim Low}.} \bibinfo{year}{2011}\natexlab{}.
\newblock \showarticletitle{{SandCanvas}: a multi-touch art medium inspired by sand animation}. In \bibinfo{booktitle}{\emph{Proceedings of the {SIGCHI} {Conference} on {Human} {Factors} in {Computing} {Systems}}} \emph{(\bibinfo{series}{{CHI} '11})}. \bibinfo{publisher}{Association for Computing Machinery}, \bibinfo{address}{New York, NY, USA}, \bibinfo{pages}{1283--1292}.
\newblock
\showISBNx{978-1-4503-0228-9}
\href{https://doi.org/10.1145/1978942.1979133}{doi:\nolinkurl{10.1145/1978942.1979133}}


\bibitem[Kim et~al\mbox{.}(2010)]%
        {kim_tablaction_2010}
\bibfield{author}{\bibinfo{person}{Hyunyoung Kim}, \bibinfo{person}{Kuenhwan Kwak}, \bibinfo{person}{Jongwoo Jung}, \bibinfo{person}{Insik Myung}, {and} \bibinfo{person}{Minsoo Hahn}.} \bibinfo{year}{2010}\natexlab{}.
\newblock \showarticletitle{Tablaction: collaborative brainstorming system with stylus-fingertip interactions on tablet {PCs}}. In \bibinfo{booktitle}{\emph{Proceedings of the 9th {ACM} {SIGGRAPH} {Conference} on {Virtual}-{Reality} {Continuum} and its {Applications} in {Industry}}}. \bibinfo{publisher}{ACM}, \bibinfo{address}{Seoul South Korea}, \bibinfo{pages}{81--84}.
\newblock
\showISBNx{978-1-4503-0459-7}
\href{https://doi.org/10.1145/1900179.1900195}{doi:\nolinkurl{10.1145/1900179.1900195}}


\bibitem[Kim et~al\mbox{.}(2021)]%
        {kim_collaborative_2021}
\bibfield{author}{\bibinfo{person}{Jingoog Kim}, \bibinfo{person}{Mary Maher}, {and} \bibinfo{person}{Safat Siddiqui}.} \bibinfo{year}{2021}\natexlab{}.
\newblock \showarticletitle{Collaborative {Ideation} {Partner}: {Design} {Ideation} in {Human}-{AI} {Co}-creativity:}. In \bibinfo{booktitle}{\emph{Proceedings of the 5th {International} {Conference} on {Computer}-{Human} {Interaction} {Research} and {Applications}}}. \bibinfo{publisher}{SCITEPRESS - Science and Technology Publications}, \bibinfo{address}{Online Streaming, --- Select a Country ---}, \bibinfo{pages}{123--130}.
\newblock
\showISBNx{978-989-758-538-8}
\href{https://doi.org/10.5220/0010640800003060}{doi:\nolinkurl{10.5220/0010640800003060}}


\bibitem[Kim et~al\mbox{.}(2024)]%
        {10.1145/3613904.3642216}
\bibfield{author}{\bibinfo{person}{Tae~Soo Kim}, \bibinfo{person}{Yoonjoo Lee}, \bibinfo{person}{Jamin Shin}, \bibinfo{person}{Young-Ho Kim}, {and} \bibinfo{person}{Juho Kim}.} \bibinfo{year}{2024}\natexlab{}.
\newblock \showarticletitle{EvalLM: Interactive Evaluation of Large Language Model Prompts on User-Defined Criteria}. In \bibinfo{booktitle}{\emph{Proceedings of the 2024 CHI Conference on Human Factors in Computing Systems}} (Honolulu, HI, USA) \emph{(\bibinfo{series}{CHI '24})}. \bibinfo{publisher}{Association for Computing Machinery}, \bibinfo{address}{New York, NY, USA}, Article \bibinfo{articleno}{306}, \bibinfo{numpages}{21}~pages.
\newblock
\showISBNx{9798400703300}
\href{https://doi.org/10.1145/3613904.3642216}{doi:\nolinkurl{10.1145/3613904.3642216}}


\bibitem[Knight et~al\mbox{.}(2021)]%
        {Knight2021IdeationGrids}
\bibfield{author}{\bibinfo{person}{John Knight}, \bibinfo{person}{Elliot Ross}, {and} \bibinfo{person}{Dan Fitton}.} \bibinfo{year}{2021}\natexlab{}.
\newblock \showarticletitle{Using Ideation Grids to Power Collaborative Creativity in Face-to-Face and Remote Innovation Sessions}.
\newblock In \bibinfo{booktitle}{\emph{Creativity - A Force to Innovation}}. \bibinfo{publisher}{IntechOpen}, \bibinfo{pages}{1--18}.
\newblock
\href{https://doi.org/10.5772/intechopen.93850}{doi:\nolinkurl{10.5772/intechopen.93850}}


\bibitem[Koch et~al\mbox{.}(2020)]%
        {koch_imagesense_2020}
\bibfield{author}{\bibinfo{person}{Janin Koch}, \bibinfo{person}{Nicolas Taffin}, \bibinfo{person}{Michel Beaudouin-Lafon}, \bibinfo{person}{Markku Laine}, \bibinfo{person}{Andrés Lucero}, {and} \bibinfo{person}{Wendy~E. Mackay}.} \bibinfo{year}{2020}\natexlab{}.
\newblock \showarticletitle{{ImageSense}: {An} {Intelligent} {Collaborative} {Ideation} {Tool} to {Support} {Diverse} {Human}-{Computer} {Partnerships}}.
\newblock \bibinfo{journal}{\emph{Proc. ACM Hum.-Comput. Interact.}} \bibinfo{volume}{4}, \bibinfo{number}{CSCW1} (\bibinfo{date}{May} \bibinfo{year}{2020}), \bibinfo{pages}{45:1--45:27}.
\newblock
\href{https://doi.org/10.1145/3392850}{doi:\nolinkurl{10.1145/3392850}}


\bibitem[Kohn et~al\mbox{.}(2011)]%
        {kohn_collaborative_2011}
\bibfield{author}{\bibinfo{person}{Nicholas Kohn}, \bibinfo{person}{Paul Paulus}, {and} \bibinfo{person}{YunHee Choi}.} \bibinfo{year}{2011}\natexlab{}.
\newblock \showarticletitle{Building on the ideas of others: An examination of the idea combination process}.
\newblock \bibinfo{journal}{\emph{Journal of Experimental Social Psychology - J EXP SOC PSYCHOL}}  \bibinfo{volume}{47} (\bibinfo{date}{05} \bibinfo{year}{2011}), \bibinfo{pages}{554--561}.
\newblock
\href{https://doi.org/10.1016/j.jesp.2011.01.004}{doi:\nolinkurl{10.1016/j.jesp.2011.01.004}}


\bibitem[Kwon et~al\mbox{.}(2024)]%
        {kwon_designer-generative_2024}
\bibfield{author}{\bibinfo{person}{Jungryun Kwon}, \bibinfo{person}{Eui-Chul Jung}, {and} \bibinfo{person}{Jiheon Kim}.} \bibinfo{year}{2024}\natexlab{}.
\newblock \showarticletitle{Designer-{Generative} {AI} {Ideation} {Process}: {Generating} {Images} {Aligned} with {Designer} {Intent} in {Early}-{Stage} {Concept} {Exploration} in {Product} {Design}}.
\newblock \bibinfo{journal}{\emph{Archives of Design Research}} \bibinfo{volume}{37}, \bibinfo{number}{3} (\bibinfo{date}{July} \bibinfo{year}{2024}), \bibinfo{pages}{7--23}.
\newblock
\showISSN{1226-8046, 2288-2987}
\href{https://doi.org/10.15187/adr.2024.07.37.3.7}{doi:\nolinkurl{10.15187/adr.2024.07.37.3.7}}


\bibitem[Langham and Paulsen(2020)]%
        {langham_simulating_2020}
\bibfield{author}{\bibinfo{person}{Jo’Anne Langham} {and} \bibinfo{person}{Neil Paulsen}.} \bibinfo{year}{2020}\natexlab{}.
\newblock \showarticletitle{Simulating the {Cognitive} {Leap} {Using} {Brainwriting}}.
\newblock \bibinfo{journal}{\emph{Design Management Journal}} \bibinfo{volume}{15}, \bibinfo{number}{1} (\bibinfo{year}{2020}), \bibinfo{pages}{85--101}.
\newblock
\showISSN{1948-7177}
\href{https://doi.org/10.1111/dmj.12062}{doi:\nolinkurl{10.1111/dmj.12062}}
\newblock
\shownote{\_eprint: https://onlinelibrary.wiley.com/doi/pdf/10.1111/dmj.12062}.


\bibitem[Lee et~al\mbox{.}(2017)]%
        {lee2017all4one}
\bibfield{author}{\bibinfo{person}{Hyelip Lee}, \bibinfo{person}{Seungwoo Je}, {and} \bibinfo{person}{Andrea Bianchi}.} \bibinfo{year}{2017}\natexlab{}.
\newblock \showarticletitle{All4One: A Moderated Sketching Tool for Supporting Idea-Generation with Remote Users}. In \bibinfo{booktitle}{\emph{Proceedings of the IASDR 2017, 7th International Conference of the International Association of Societies of Design Research}}. \bibinfo{publisher}{International Association of Societies of Design Research}, \bibinfo{address}{Cincinnati, USA}.
\newblock
\href{https://doi.org/10.7945/C2JH5D}{doi:\nolinkurl{10.7945/C2JH5D}}


\bibitem[Lin et~al\mbox{.}(2025)]%
        {lin_inkspire_2025}
\bibfield{author}{\bibinfo{person}{David Chuan-En Lin}, \bibinfo{person}{Hyeonsu~B. Kang}, \bibinfo{person}{Nikolas Martelaro}, \bibinfo{person}{Aniket Kittur}, \bibinfo{person}{Yan-Ying Chen}, {and} \bibinfo{person}{Matthew~K. Hong}.} \bibinfo{year}{2025}\natexlab{}.
\newblock \showarticletitle{Inkspire: {Supporting} {Design} {Exploration} with {Generative} {AI} through {Analogical} {Sketching}}. In \bibinfo{booktitle}{\emph{Proceedings of the 2025 {CHI} {Conference} on {Human} {Factors} in {Computing} {Systems}}}. \bibinfo{publisher}{ACM}, \bibinfo{address}{Yokohama Japan}, \bibinfo{pages}{1--18}.
\newblock
\showISBNx{979-8-4007-1394-1}
\href{https://doi.org/10.1145/3706598.3713397}{doi:\nolinkurl{10.1145/3706598.3713397}}


\bibitem[{McHugh}(2012)]%
        {mchugh_interrater_2012}
\bibfield{author}{\bibinfo{person}{Mary~L. {McHugh}}.} \bibinfo{year}{2012}\natexlab{}.
\newblock \showarticletitle{Interrater reliability: the kappa statistic}.
\newblock  \bibinfo{volume}{22}, \bibinfo{number}{3} (\bibinfo{year}{2012}), \bibinfo{pages}{276--282}.
\newblock
\showISSN{1330-0962}


\bibitem[Melniczuk et~al\mbox{.}(2024)]%
        {10.1145/3623509.3633361}
\bibfield{author}{\bibinfo{person}{Amy Melniczuk}, \bibinfo{person}{Nam~Gu Kang}, {and} \bibinfo{person}{Shaun Lawson}.} \bibinfo{year}{2024}\natexlab{}.
\newblock \showarticletitle{Embedding Thinking Strategies within a Tangible Tree to Orchestrate Small Group Brainstorming}. In \bibinfo{booktitle}{\emph{Proceedings of the Eighteenth International Conference on Tangible, Embedded, and Embodied Interaction}} (Cork, Ireland) \emph{(\bibinfo{series}{TEI '24})}. \bibinfo{publisher}{Association for Computing Machinery}, \bibinfo{address}{New York, NY, USA}, Article \bibinfo{articleno}{13}, \bibinfo{numpages}{18}~pages.
\newblock
\showISBNx{9798400704024}
\href{https://doi.org/10.1145/3623509.3633361}{doi:\nolinkurl{10.1145/3623509.3633361}}


\bibitem[Melniczuk et~al\mbox{.}(2025)]%
        {melniczuk_exploring_2025}
\bibfield{author}{\bibinfo{person}{Amy Melniczuk}, \bibinfo{person}{Meng Liang}, {and} \bibinfo{person}{Julian Preissing}.} \bibinfo{year}{2025}\natexlab{}.
\newblock \showarticletitle{Exploring {Tangible} {Designs} to {Improve} {Interpersonal} {Connectedness} in {Remote} {Group} {Brainstorming}}. In \bibinfo{booktitle}{\emph{Proceedings of the {Nineteenth} {International} {Conference} on {Tangible}, {Embedded}, and {Embodied} {Interaction}}} \emph{(\bibinfo{series}{{TEI} '25})}. \bibinfo{publisher}{Association for Computing Machinery}, \bibinfo{address}{New York, NY, USA}, \bibinfo{pages}{1--14}.
\newblock
\showISBNx{979-8-4007-1197-8}
\href{https://doi.org/10.1145/3689050.3704933}{doi:\nolinkurl{10.1145/3689050.3704933}}


\bibitem[Nijstad and Stroebe(2006)]%
        {nijstad_production_2002}
\bibfield{author}{\bibinfo{person}{Bernard~A. Nijstad} {and} \bibinfo{person}{Wolfgang Stroebe}.} \bibinfo{year}{2006}\natexlab{}.
\newblock \showarticletitle{How the Group Affects the Mind: A Cognitive Model of Idea Generation in Groups}.
\newblock \bibinfo{journal}{\emph{Personality and Social Psychology Review}} \bibinfo{volume}{10}, \bibinfo{number}{3} (\bibinfo{year}{2006}), \bibinfo{pages}{186--213}.
\newblock
\href{https://doi.org/10.1207/s15327957pspr1003\_1}{doi:\nolinkurl{10.1207/s15327957pspr1003\_1}}
\newblock
\shownote{PMID: 16859437}.


\bibitem[Pan et~al\mbox{.}(2024)]%
        {pan2024humancentereddesignrecommendationsllmasajudge}
\bibfield{author}{\bibinfo{person}{Qian Pan}, \bibinfo{person}{Zahra Ashktorab}, \bibinfo{person}{Michael Desmond}, \bibinfo{person}{Martin~Santillan Cooper}, \bibinfo{person}{James Johnson}, \bibinfo{person}{Rahul Nair}, \bibinfo{person}{Elizabeth Daly}, {and} \bibinfo{person}{Werner Geyer}.} \bibinfo{year}{2024}\natexlab{}.
\newblock \bibinfo{title}{Human-Centered Design Recommendations for LLM-as-a-Judge}.
\newblock
\showeprint[arxiv]{2407.03479}~[cs.HC]
\urldef\tempurl%
\url{https://arxiv.org/abs/2407.03479}
\showURL{%
\tempurl}


\bibitem[Rombach et~al\mbox{.}(2022)]%
        {rombach_high-resolution_2022}
\bibfield{author}{\bibinfo{person}{Robin Rombach}, \bibinfo{person}{Andreas Blattmann}, \bibinfo{person}{Dominik Lorenz}, \bibinfo{person}{Patrick Esser}, {and} \bibinfo{person}{Björn Ommer}.} \bibinfo{year}{2022}\natexlab{}.
\newblock \bibinfo{title}{High-{Resolution} {Image} {Synthesis} with {Latent} {Diffusion} {Models}}.
\newblock
\href{https://doi.org/10.48550/arXiv.2112.10752}{doi:\nolinkurl{10.48550/arXiv.2112.10752}}
\newblock
\shownote{arXiv:2112.10752 [cs]}.


\bibitem[Runco and Acar(2012)]%
        {runco_divergent_2012}
\bibfield{author}{\bibinfo{person}{Mark~A. Runco} {and} \bibinfo{person}{Selcuk Acar}.} \bibinfo{year}{2012}\natexlab{}.
\newblock \showarticletitle{Divergent Thinking as an Indicator of Creative Potential}.
\newblock \bibinfo{journal}{\emph{Creativity Research Journal}} \bibinfo{volume}{24}, \bibinfo{number}{1} (\bibinfo{year}{2012}), \bibinfo{pages}{66--75}.
\newblock
\href{https://doi.org/10.1080/10400419.2012.652929}{doi:\nolinkurl{10.1080/10400419.2012.652929}}


\bibitem[Saad et~al\mbox{.}(2023)]%
        {saad_sensing_2023}
\bibfield{author}{\bibinfo{person}{Alia Saad}, \bibinfo{person}{Kian Izadi}, \bibinfo{person}{Anam Ahmad~Khan}, \bibinfo{person}{Pascal Knierim}, \bibinfo{person}{Stefan Schneegass}, \bibinfo{person}{Florian Alt}, {and} \bibinfo{person}{Yomna Abdelrahman}.} \bibinfo{year}{2023}\natexlab{}.
\newblock \showarticletitle{HotFoot: Foot-Based User Identification Using Thermal Imaging}. In \bibinfo{booktitle}{\emph{Proceedings of the 2023 CHI Conference on Human Factors in Computing Systems}} (Hamburg, Germany) \emph{(\bibinfo{series}{CHI '23})}. \bibinfo{publisher}{Association for Computing Machinery}, \bibinfo{address}{New York, NY, USA}, Article \bibinfo{articleno}{262}, \bibinfo{numpages}{13}~pages.
\newblock
\showISBNx{9781450394215}
\href{https://doi.org/10.1145/3544548.3580924}{doi:\nolinkurl{10.1145/3544548.3580924}}


\bibitem[Saharia et~al\mbox{.}(2022)]%
        {saharia_photorealistic_2022}
\bibfield{author}{\bibinfo{person}{Chitwan Saharia}, \bibinfo{person}{William Chan}, \bibinfo{person}{Saurabh Saxena}, \bibinfo{person}{Lala Li}, \bibinfo{person}{Jay Whang}, \bibinfo{person}{Emily Denton}, \bibinfo{person}{Seyed Kamyar~Seyed Ghasemipour}, \bibinfo{person}{Burcu~Karagol Ayan}, \bibinfo{person}{S.~Sara Mahdavi}, \bibinfo{person}{Rapha~Gontijo Lopes}, \bibinfo{person}{Tim Salimans}, \bibinfo{person}{Jonathan Ho}, \bibinfo{person}{David~J. Fleet}, {and} \bibinfo{person}{Mohammad Norouzi}.} \bibinfo{year}{2022}\natexlab{}.
\newblock \bibinfo{title}{Photorealistic {Text}-to-{Image} {Diffusion} {Models} with {Deep} {Language} {Understanding}}.
\newblock
\href{https://doi.org/10.48550/arXiv.2205.11487}{doi:\nolinkurl{10.48550/arXiv.2205.11487}}
\newblock
\shownote{arXiv:2205.11487 [cs]}.


\bibitem[Sch{\"o}n(1992)]%
        {schon1992reflective}
\bibfield{author}{\bibinfo{person}{Donald~A. Sch{\"o}n}.} \bibinfo{year}{1992}\natexlab{}.
\newblock \bibinfo{booktitle}{\emph{The Reflective Practitioner: How Professionals Think in Action} (\bibinfo{edition}{1} ed.)}.
\newblock \bibinfo{publisher}{Routledge}.
\newblock
\href{https://doi.org/10.4324/9781315237473}{doi:\nolinkurl{10.4324/9781315237473}}


\bibitem[Shaer et~al\mbox{.}(2024)]%
        {shaer_ai-augmented_2024}
\bibfield{author}{\bibinfo{person}{Orit Shaer}, \bibinfo{person}{Angelora Cooper}, \bibinfo{person}{Osnat Mokryn}, \bibinfo{person}{Andrew~L Kun}, {and} \bibinfo{person}{Hagit Ben~Shoshan}.} \bibinfo{year}{2024}\natexlab{}.
\newblock \showarticletitle{{AI}-{Augmented} {Brainwriting}: {Investigating} the use of {LLMs} in group ideation}. In \bibinfo{booktitle}{\emph{Proceedings of the 2024 {CHI} {Conference} on {Human} {Factors} in {Computing} {Systems}}} \emph{(\bibinfo{series}{{CHI} '24})}. \bibinfo{publisher}{Association for Computing Machinery}, \bibinfo{address}{New York, NY, USA}, \bibinfo{pages}{1--17}.
\newblock
\showISBNx{979-8-4007-0330-0}
\href{https://doi.org/10.1145/3613904.3642414}{doi:\nolinkurl{10.1145/3613904.3642414}}


\bibitem[Shah et~al\mbox{.}(2001)]%
        {shah2001collaborative}
\bibfield{author}{\bibinfo{person}{Jami~J. Shah}, \bibinfo{person}{Noel Vargas-Hernandez}, \bibinfo{person}{Joshua~D. Summers}, {and} \bibinfo{person}{Sanjay Kulkarni}.} \bibinfo{year}{2001}\natexlab{}.
\newblock \showarticletitle{Collaborative sketching (c-sketch)—An idea generation technique for engineering design}.
\newblock \bibinfo{journal}{\emph{The Journal of Creative Behavior}} \bibinfo{volume}{35}, \bibinfo{number}{3} (\bibinfo{year}{2001}), \bibinfo{pages}{168--198}.
\newblock
\href{https://doi.org/10.1002/j.2162-6057.2001.tb01045.x}{doi:\nolinkurl{10.1002/j.2162-6057.2001.tb01045.x}}


\bibitem[Shin et~al\mbox{.}(2025)]%
        {Joongi2025}
\bibfield{author}{\bibinfo{person}{Joongi Shin}, \bibinfo{person}{Ankit Khatri}, \bibinfo{person}{Michael~A. Hedderich}, \bibinfo{person}{Andr\'{e}s Lucero}, {and} \bibinfo{person}{Antti Oulasvirta}.} \bibinfo{year}{2025}\natexlab{}.
\newblock \showarticletitle{Facilitating Asynchronous Idea Generation and Selection with Chatbots}. In \bibinfo{booktitle}{\emph{Proceedings of the 36th Australasian Conference on Human-Computer Interaction}} \emph{(\bibinfo{series}{OzCHI '24})}. \bibinfo{publisher}{Association for Computing Machinery}, \bibinfo{address}{New York, NY, USA}, \bibinfo{pages}{305–323}.
\newblock
\showISBNx{9798400715099}
\href{https://doi.org/10.1145/3726986.3726994}{doi:\nolinkurl{10.1145/3726986.3726994}}


\bibitem[Shneiderman(1983)]%
        {shneiderman_direct_1983}
\bibfield{author}{\bibinfo{person}{Ben Shneiderman}.} \bibinfo{year}{1983}\natexlab{}.
\newblock \showarticletitle{Direct manipulation: A step beyond programming languages}.
\newblock \bibinfo{journal}{\emph{Computer}} \bibinfo{volume}{16}, \bibinfo{number}{8} (\bibinfo{year}{1983}), \bibinfo{pages}{57--69}.
\newblock
\href{https://doi.org/10.1109/MC.1983.1654471}{doi:\nolinkurl{10.1109/MC.1983.1654471}}


\bibitem[Shneiderman(2007)]%
        {shneiderman_creativity_2007}
\bibfield{author}{\bibinfo{person}{Ben Shneiderman}.} \bibinfo{year}{2007}\natexlab{}.
\newblock \showarticletitle{Creativity support tools: accelerating discovery and innovation}.
\newblock \bibinfo{journal}{\emph{Commun. ACM}} \bibinfo{volume}{50}, \bibinfo{number}{12} (\bibinfo{date}{Dec.} \bibinfo{year}{2007}), \bibinfo{pages}{20--32}.
\newblock
\showISSN{0001-0782}
\href{https://doi.org/10.1145/1323688.1323689}{doi:\nolinkurl{10.1145/1323688.1323689}}


\bibitem[Siangliulue et~al\mbox{.}(2016)]%
        {siangliulue_ideahound_2016}
\bibfield{author}{\bibinfo{person}{Pao Siangliulue}, \bibinfo{person}{Joel Chan}, \bibinfo{person}{Steven~P. Dow}, {and} \bibinfo{person}{Krzysztof~Z. Gajos}.} \bibinfo{year}{2016}\natexlab{}.
\newblock \showarticletitle{{IdeaHound}: {Improving} {Large}-scale {Collaborative} {Ideation} with {Crowd}-{Powered} {Real}-time {Semantic} {Modeling}}. In \bibinfo{booktitle}{\emph{Proceedings of the 29th {Annual} {Symposium} on {User} {Interface} {Software} and {Technology}}} \emph{(\bibinfo{series}{{UIST} '16})}. \bibinfo{publisher}{Association for Computing Machinery}, \bibinfo{address}{New York, NY, USA}, \bibinfo{pages}{609--624}.
\newblock
\showISBNx{978-1-4503-4189-9}
\href{https://doi.org/10.1145/2984511.2984578}{doi:\nolinkurl{10.1145/2984511.2984578}}


\bibitem[Son et~al\mbox{.}(2024)]%
        {10.1145/3613904.3642847}
\bibfield{author}{\bibinfo{person}{Kihoon Son}, \bibinfo{person}{DaEun Choi}, \bibinfo{person}{Tae~Soo Kim}, \bibinfo{person}{Young-Ho Kim}, {and} \bibinfo{person}{Juho Kim}.} \bibinfo{year}{2024}\natexlab{}.
\newblock \showarticletitle{GenQuery: Supporting Expressive Visual Search with Generative Models}. In \bibinfo{booktitle}{\emph{Proceedings of the 2024 CHI Conference on Human Factors in Computing Systems}} (Honolulu, HI, USA) \emph{(\bibinfo{series}{CHI '24})}. \bibinfo{publisher}{Association for Computing Machinery}, \bibinfo{address}{New York, NY, USA}, Article \bibinfo{articleno}{180}, \bibinfo{numpages}{19}~pages.
\newblock
\showISBNx{9798400703300}
\href{https://doi.org/10.1145/3613904.3642847}{doi:\nolinkurl{10.1145/3613904.3642847}}


\bibitem[Song et~al\mbox{.}(2025)]%
        {Jiaxin_2025}
\bibfield{author}{\bibinfo{person}{Jiaxin Song}, \bibinfo{person}{Parnian Shahkar}, \bibinfo{person}{Kate Donahue}, {and} \bibinfo{person}{Bhaskar~Ray Chaudhury}.} \bibinfo{year}{2025}\natexlab{}.
\newblock \showarticletitle{Human-AI Collaboration with Misaligned Preferences}. In \bibinfo{booktitle}{\emph{Proceedings of the 5th ACM Conference on Equity and Access in Algorithms, Mechanisms, and Optimization}} \emph{(\bibinfo{series}{EAAMO '25})}. \bibinfo{publisher}{Association for Computing Machinery}, \bibinfo{address}{New York, NY, USA}, \bibinfo{pages}{294}.
\newblock
\showISBNx{9798400721403}
\href{https://doi.org/10.1145/3757887.3767691}{doi:\nolinkurl{10.1145/3757887.3767691}}


\bibitem[Sun et~al\mbox{.}(2025)]%
        {sun_creative_2025}
\bibfield{author}{\bibinfo{person}{Zhida Sun}, \bibinfo{person}{Zhenyao Zhang}, \bibinfo{person}{Yue Zhang}, \bibinfo{person}{Min Lu}, \bibinfo{person}{Dani Lischinski}, \bibinfo{person}{Daniel Cohen-Or}, {and} \bibinfo{person}{Hui Huang}.} \bibinfo{year}{2025}\natexlab{}.
\newblock \bibinfo{title}{Creative {Blends} of {Visual} {Concepts}}.
\newblock
\href{https://doi.org/10.48550/arXiv.2502.16062}{doi:\nolinkurl{10.48550/arXiv.2502.16062}}
\newblock
\shownote{arXiv:2502.16062 [cs]}.


\bibitem[Sutton and Hargadon(1996)]%
        {692e1dfc-42e1-3d68-a556-1bc77d50cbd3}
\bibfield{author}{\bibinfo{person}{Robert~I. Sutton} {and} \bibinfo{person}{Andrew Hargadon}.} \bibinfo{year}{1996}\natexlab{}.
\newblock \showarticletitle{Brainstorming Groups in Context: Effectiveness in a Product Design Firm}.
\newblock \bibinfo{journal}{\emph{Administrative Science Quarterly}} \bibinfo{volume}{41}, \bibinfo{number}{4} (\bibinfo{year}{1996}), \bibinfo{pages}{685--718}.
\newblock
\showISSN{00018392}
\urldef\tempurl%
\url{http://www.jstor.org/stable/2393872}
\showURL{%
\tempurl}


\bibitem[Suwa and Tversky(1997)]%
        {suwa_what_1997}
\bibfield{author}{\bibinfo{person}{Masaki Suwa} {and} \bibinfo{person}{Barbara Tversky}.} \bibinfo{year}{1997}\natexlab{}.
\newblock \showarticletitle{What do architects and students perceive in their design sketches? {A} protocol analysis}.
\newblock \bibinfo{journal}{\emph{Design Studies}} \bibinfo{volume}{18}, \bibinfo{number}{4} (\bibinfo{date}{Oct.} \bibinfo{year}{1997}), \bibinfo{pages}{385--403}.
\newblock
\showISSN{0142-694X}
\href{https://doi.org/10.1016/S0142-694X(97)00008-2}{doi:\nolinkurl{10.1016/S0142-694X(97)00008-2}}


\bibitem[Thomas(2006)]%
        {thomas_general_2006}
\bibfield{author}{\bibinfo{person}{David~R. Thomas}.} \bibinfo{year}{2006}\natexlab{}.
\newblock \showarticletitle{A general inductive approach for analyzing qualitative evaluation data}.
\newblock  \bibinfo{volume}{27}, \bibinfo{number}{2} (\bibinfo{year}{2006}), \bibinfo{pages}{237--246}.
\newblock
\showISSN{1557-0878}
\href{https://doi.org/10.1177/1098214005283748}{doi:\nolinkurl{10.1177/1098214005283748}}
\newblock
\shownote{Place: {US} Publisher: Sage Publications}.


\bibitem[Tversky(2011)]%
        {tversky_visualizing_2011}
\bibfield{author}{\bibinfo{person}{Barbara Tversky}.} \bibinfo{year}{2011}\natexlab{}.
\newblock \showarticletitle{Visualizing {Thought}}.
\newblock \bibinfo{journal}{\emph{Topics in Cognitive Science}} \bibinfo{volume}{3}, \bibinfo{number}{3} (\bibinfo{year}{2011}), \bibinfo{pages}{499--535}.
\newblock
\showISSN{1756-8765}
\href{https://doi.org/10.1111/j.1756-8765.2010.01113.x}{doi:\nolinkurl{10.1111/j.1756-8765.2010.01113.x}}
\newblock
\shownote{\_eprint: https://onlinelibrary.wiley.com/doi/pdf/10.1111/j.1756-8765.2010.01113.x}.


\bibitem[Van Der~Lugt(2002)]%
        {van_der_lugt_brainsketching_2002}
\bibfield{author}{\bibinfo{person}{Remko Van Der~Lugt}.} \bibinfo{year}{2002}\natexlab{}.
\newblock \showarticletitle{Brainsketching and {How} it {Differs} from {Brainstorming}}.
\newblock \bibinfo{journal}{\emph{Creativity and Innovation Management}} \bibinfo{volume}{11}, \bibinfo{number}{1} (\bibinfo{year}{2002}), \bibinfo{pages}{43--54}.
\newblock
\showISSN{1467-8691}
\href{https://doi.org/10.1111/1467-8691.00235}{doi:\nolinkurl{10.1111/1467-8691.00235}}
\newblock
\shownote{\_eprint: https://onlinelibrary.wiley.com/doi/pdf/10.1111/1467-8691.00235}.


\bibitem[Wan and Lu(2023)]%
        {wan_gancollage_2023}
\bibfield{author}{\bibinfo{person}{Qian Wan} {and} \bibinfo{person}{Zhicong Lu}.} \bibinfo{year}{2023}\natexlab{}.
\newblock \showarticletitle{{GANCollage}: {A} {GAN}-{Driven} {Digital} {Mood} {Board} to {Facilitate} {Ideation} in {Creativity} {Support}}. In \bibinfo{booktitle}{\emph{Proceedings of the 2023 {ACM} {Designing} {Interactive} {Systems} {Conference}}} \emph{(\bibinfo{series}{{DIS} '23})}. \bibinfo{publisher}{Association for Computing Machinery}, \bibinfo{address}{New York, NY, USA}, \bibinfo{pages}{136--146}.
\newblock
\showISBNx{978-1-4503-9893-0}
\href{https://doi.org/10.1145/3563657.3596072}{doi:\nolinkurl{10.1145/3563657.3596072}}


\bibitem[Wang et~al\mbox{.}(2023)]%
        {wang_popblends_2023}
\bibfield{author}{\bibinfo{person}{Sitong Wang}, \bibinfo{person}{Savvas Petridis}, \bibinfo{person}{Taeahn Kwon}, \bibinfo{person}{Xiaojuan Ma}, {and} \bibinfo{person}{Lydia~B Chilton}.} \bibinfo{year}{2023}\natexlab{}.
\newblock \showarticletitle{{PopBlends}: {Strategies} for {Conceptual} {Blending} with {Large} {Language} {Models}}. In \bibinfo{booktitle}{\emph{Proceedings of the 2023 {CHI} {Conference} on {Human} {Factors} in {Computing} {Systems}}} \emph{(\bibinfo{series}{{CHI} '23})}. \bibinfo{publisher}{Association for Computing Machinery}, \bibinfo{address}{New York, NY, USA}, \bibinfo{pages}{1--19}.
\newblock
\showISBNx{978-1-4503-9421-5}
\href{https://doi.org/10.1145/3544548.3580948}{doi:\nolinkurl{10.1145/3544548.3580948}}


\bibitem[Wang et~al\mbox{.}(2025)]%
        {wang_aideation_2025}
\bibfield{author}{\bibinfo{person}{Wen-Fan Wang}, \bibinfo{person}{Chien-Ting Lu}, \bibinfo{person}{Nil Ponsa~i Campanyà}, \bibinfo{person}{Bing-Yu Chen}, {and} \bibinfo{person}{Mike~Y. Chen}.} \bibinfo{year}{2025}\natexlab{}.
\newblock \showarticletitle{{AIdeation}: {Designing} a {Human}-{AI} {Collaborative} {Ideation} {System} for {Concept} {Designers}}. In \bibinfo{booktitle}{\emph{Proceedings of the 2025 {CHI} {Conference} on {Human} {Factors} in {Computing} {Systems}}} \emph{(\bibinfo{series}{{CHI} '25})}. \bibinfo{publisher}{Association for Computing Machinery}, \bibinfo{address}{New York, NY, USA}, \bibinfo{pages}{1--28}.
\newblock
\showISBNx{979-8-4007-1394-1}
\href{https://doi.org/10.1145/3706598.3714148}{doi:\nolinkurl{10.1145/3706598.3714148}}


\bibitem[Webb(2015)]%
        {Webb2015PenTouchDiagramming}
\bibfield{author}{\bibinfo{person}{Andrew~M. Webb}.} \bibinfo{year}{2015}\natexlab{}.
\newblock \showarticletitle{Pen + Touch Diagramming to Stimulate Design Ideation}. In \bibinfo{booktitle}{\emph{Proceedings of the ACM SIGCHI Conference on Creativity and Cognition}}. \bibinfo{pages}{331--332}.
\newblock
\href{https://doi.org/10.1145/2757226.2764766}{doi:\nolinkurl{10.1145/2757226.2764766}}
\newblock
\shownote{C\&C '15}.


\bibitem[Wilson(2013)]%
        {wilson_chapter_2013}
\bibfield{author}{\bibinfo{person}{Chauncey Wilson}.} \bibinfo{year}{2013}\natexlab{}.
\newblock \showarticletitle{Chapter 2 - Brainwriting}.
\newblock In \bibinfo{booktitle}{\emph{Brainstorming and Beyond}}, \bibfield{editor}{\bibinfo{person}{Chauncey Wilson}} (Ed.). \bibinfo{publisher}{Morgan Kaufmann}, \bibinfo{pages}{43--61}.
\newblock
\showISBNx{978-0-12-407157-5}
\href{https://doi.org/10.1016/B978-0-12-407157-5.00002-6}{doi:\nolinkurl{10.1016/B978-0-12-407157-5.00002-6}}


\bibitem[Zhang et~al\mbox{.}(2023)]%
        {zhang_generative_2023}
\bibfield{author}{\bibinfo{person}{Chengzhi Zhang}, \bibinfo{person}{Weijie Wang}, \bibinfo{person}{Paul Pangaro}, \bibinfo{person}{Nikolas Martelaro}, {and} \bibinfo{person}{Daragh Byrne}.} \bibinfo{year}{2023}\natexlab{}.
\newblock \showarticletitle{Generative {Image} {AI} {Using} {Design} {Sketches} as input: {Opportunities} and {Challenges}}. In \bibinfo{booktitle}{\emph{Proceedings of the 15th {Conference} on {Creativity} and {Cognition}}} \emph{(\bibinfo{series}{C\&{C} '23})}. \bibinfo{publisher}{Association for Computing Machinery}, \bibinfo{address}{New York, NY, USA}, \bibinfo{pages}{254--261}.
\newblock
\showISBNx{979-8-4007-0180-1}
\href{https://doi.org/10.1145/3591196.3596820}{doi:\nolinkurl{10.1145/3591196.3596820}}


\bibitem[Zhang et~al\mbox{.}(2025)]%
        {zhang_ladica_2025}
\bibfield{author}{\bibinfo{person}{Zheng Zhang}, \bibinfo{person}{Weirui Peng}, \bibinfo{person}{Xinyue Chen}, \bibinfo{person}{Luke Cao}, {and} \bibinfo{person}{Toby Jia-Jun Li}.} \bibinfo{year}{2025}\natexlab{}.
\newblock \showarticletitle{{LADICA}: {A} {Large} {Shared} {Display} {Interface} for {Generative} {AI} {Cognitive} {Assistance} in {Co}-located {Team} {Collaboration}}. In \bibinfo{booktitle}{\emph{Proceedings of the 2025 {CHI} {Conference} on {Human} {Factors} in {Computing} {Systems}}} \emph{(\bibinfo{series}{{CHI} '25})}. \bibinfo{publisher}{Association for Computing Machinery}, \bibinfo{address}{New York, NY, USA}, \bibinfo{pages}{1--22}.
\newblock
\showISBNx{979-8-4007-1394-1}
\href{https://doi.org/10.1145/3706598.3713289}{doi:\nolinkurl{10.1145/3706598.3713289}}


\bibitem[Zhao et~al\mbox{.}(2014)]%
        {zhao_skwiki_2014}
\bibfield{author}{\bibinfo{person}{Zhenpeng Zhao}, \bibinfo{person}{Sriram~Karthik Badam}, \bibinfo{person}{Senthil Chandrasegaran}, \bibinfo{person}{Deok~Gun Park}, \bibinfo{person}{Niklas Elmqvist}, \bibinfo{person}{Lorraine Kisselburgh}, {and} \bibinfo{person}{Karthik Ramani}.} \bibinfo{year}{2014}\natexlab{}.
\newblock \showarticletitle{{skWiki}: {A} {Multimedia} {Sketching} {System} for {Collaborative} {Creativity}}. In \bibinfo{booktitle}{\emph{Proceedings of the {SIGCHI} {Conference} on {Human} {Factors} in {Computing} {Systems}}}. \bibinfo{pages}{1235--1244}.
\newblock
\href{https://doi.org/10.1145/2556288.2557394}{doi:\nolinkurl{10.1145/2556288.2557394}}


\bibitem[Zhou et~al\mbox{.}(2023)]%
        {zhou2023leasttomostpromptingenablescomplex}
\bibfield{author}{\bibinfo{person}{Denny Zhou}, \bibinfo{person}{Nathanael Schärli}, \bibinfo{person}{Le Hou}, \bibinfo{person}{Jason Wei}, \bibinfo{person}{Nathan Scales}, \bibinfo{person}{Xuezhi Wang}, \bibinfo{person}{Dale Schuurmans}, \bibinfo{person}{Claire Cui}, \bibinfo{person}{Olivier Bousquet}, \bibinfo{person}{Quoc Le}, {and} \bibinfo{person}{Ed Chi}.} \bibinfo{year}{2023}\natexlab{}.
\newblock \bibinfo{title}{Least-to-Most Prompting Enables Complex Reasoning in Large Language Models}.
\newblock
\showeprint[arxiv]{2205.10625}~[cs.AI]
\urldef\tempurl%
\url{https://arxiv.org/abs/2205.10625}
\showURL{%
\tempurl}


\end{thebibliography}
